\documentclass{aa}

\usepackage{graphicx}
\usepackage{txfonts}
\usepackage[colorlinks,
  linkcolor={red!50!black},
  citecolor={green!50!black},
  urlcolor={blue!50!black}]{hyperref}

\usepackage{microtype}
\usepackage{amsmath}
\usepackage{subcaption}
\usepackage{algorithm2e}
\usepackage{multirow}
\usepackage{listings}

\usepackage{pgfplots}
\usepackage{pgfplotstable}
\usepgfplotslibrary{groupplots}

\usetikzlibrary{calc}
\usetikzlibrary{patterns}
\usetikzlibrary{arrows.meta}
\usetikzlibrary{positioning}
\usetikzlibrary{matrix}
\usetikzlibrary{external}
\usetikzlibrary{shapes}
\pgfplotsset{compat=1.13}
\pgfplotsset{every tick label/.append style={font=\tiny}}

\bibpunct{(}{)}{;}{a}{}{,} 
\DeclareFontFamily{U}{wncy}{}
\DeclareFontShape{U}{wncy}{m}{n}{<->wncyr10}{}
\DeclareSymbolFont{mcy}{U}{wncy}{m}{n}
\DeclareMathSymbol{\Sh}{\mathord}{mcy}{"58}

\newcommand{\xx}{u}
\newcommand{\xxx}{v}
\newcommand{\yy}{l}
\newcommand{\yyy}{m}
\newcommand{\xf}{\xx_{\mathrm f}}
\newcommand{\yf}{\yy_{\mathrm f}}
\newcommand{\smpx}[1]{\Sh_{#1}}
\newcommand{\smpy}[1]{\Sh_{(#1)^{\shortminus1}}}

\newcommand{\xV}{\xx_{\V}}
\newcommand{\yV}{\yy_{\V}}
\newcommand{\xA}{\xx_{\A}}
\newcommand{\yB}{\yy_{\B}}
\newcommand{\yP}{\yy_{\n\B}}
\newcommand{\xn}{\xx_{\n}}
\newcommand{\xg}{\xx_{\g_w}}
\newcommand{\xgn}{\xx_{\g_w\n}}
\newcommand{\yn}{\yy_{\n}}
\newcommand{\xm}{\xx_{\m}}

\newcommand{\A}{\mathrm{A}}
\newcommand{\B}{\mathrm{B}}
\newcommand{\V}{\Upsilon}
\newcommand{\m}{\mathrm{m}}

\newcommand{\n}{\mathrm{n}}
\newcommand{\nb}{\mathrm{b}}
\newcommand{\g}{\mathrm{g}}

\newcommand{\uu}{^\text{(u)}}

\newcommand{\uj}{\uu_j}
\newcommand{\ui}{^\text{(u)}_i}
\newcommand{\vj}{^\text{(v)}_j}
\newcommand{\vi}{^\text{(v)}_i}

\newcommand{\dxi}{\hat\xx_i}
\newcommand{\dxxi}{\hat\xxx_i}
\newcommand{\dwi}{\hat w_i}
\newcommand{\twi}{\bar w_i}
\newcommand{\dyj}{\hat\yy_j}
\newcommand{\dyyj}{\hat\yyy_j}
\newcommand{\si}{\delta_{\shortminus\dxi}}
\newcommand{\sii}{\delta_{\dxi}}

\newcommand{\tj}{\mathcal F^{\shortminus1}\!\delta_{\shortminus\dyj}}
\newcommand{\tji}{\mathcal F^{\shortminus1}\!\delta_{\dyj}}

\DeclareMathSymbol{\shortminus}{\mathbin}{AMSa}{"39}
\begin{document}

\renewcommand*{\equationautorefname}{Eq.}
\renewcommand*{\figureautorefname}{Fig.}
\renewcommand*{\sectionautorefname}{Sect.}
\renewcommand*{\subsectionautorefname}{Sect.}
\renewcommand*{\algorithmautorefname}{Alg.}
\renewcommand*{\algorithmcfname}{Alg.}
\SetAlgoCaptionLayout{centerline}
\newcommand{\mycapsty}[1]{#1}
\SetAlCapSty{mycapsty}

\title{Imaging swiFTly: streaming widefield \\ Fourier Transforms
  for large-scale interferometry}

\author{
  P. Wortmann \and J. Kent \and B. Nikolic }

\institute{Cavendish Astrophysics Group, University of Cambridge,
  JJ Thomson Avenue, Cambridge CB3 0HE, UK\\
  \email{
    peter.wortmann@skao.int,
    jameschristopherkent@gmail.com,
    bn204@cam.ac.uk}}

\date{Received ???; accepted ???}

  \abstract
    {}          { We describe a scalable distributed imaging algorithm framework for next-generation
     radio telescopes, managing the Fourier transform from
     apertures to sky (or vice versa) with a focus on minimising memory load,
     data transfers, and computation.
     }
     { Our algorithm uses
     smooth
     window functions to isolate the influence between
     specific
     regions of spatial-frequency and image space. This allows the distribution of
     image data between nodes and the construction of segments of frequency space
     exactly when and where needed.
          }
      { The developed prototype distributes terabytes of
   image data across many nodes, while generating
   visibilities at throughput and accuracy
   competitive with existing software.
   Scaling is demonstrated to be better than cubic in
   problem complexity (for baseline length and field of view), reducing
   the risk involved in growing radio astronomy processing to
   large telescopes like the Square Kilometre Array.
   }
     {}

   \keywords{Methods: numerical, Methods: data analysis, Techniques:
     interferometric, Techniques: image processing}

   \maketitle

\section{Introduction}

Upcoming large-scale radio telescopes like the Square Kilometre
Array (SKA) \,\citep{dewdney2009square} are designed for deep observations of large
areas of the sky.  These observations require many long baselines, wide
bandwidths, and high frequency resolution. This translates to demanding
computing requirements; for SKA imaging observations, more than a petabyte of
visibility data might be generated per hour. Pipelines will need to perform approximately 40 exa-operations to produce images that are terabytes in
size\,\citep{SDPParamModel2016}.  This means that efficient use of
high-performance computing facilities will be essential, which will require
algorithms that distribute computation, visibility data, as well as image data
evenly across many compute nodes.

This distribution is not trivial because radio astronomy imaging requires a
Fourier transform where every measured visibility impacts every image
pixel.  Conventional strategies for distributing radio-interferometric imaging
split visibilities by frequency or by observation time, but replicate full copies of the image to every node.  A well-known
alternative approach is to instead split the image into `facets', previously used for
non-coplanarity\,\citep{cornwell1992radio}, direction-dependent calibration,
and deconvolution\,\citep{van2016lofar}. This requires visibilities to be
gridded to a separate low-resolution grid for every facet, which combined with
averaging can be a viable scaling
strategy\,\citep{SDPParamModel2016,tasse2018faceting}.  On the other hand,
repeated gridding is inefficient, phase rotation can become a bottleneck, and
averaging introduces inherent inaccuracies into the result.


\begin{figure}[t]
    \centering
  \begin{tikzpicture}[yscale=1.2, xscale=1, inner sep=0]

    \node[above] at (7.5,2) { \begin{tabular}{c} Facets \\ \small (held) \end{tabular} };
    \node[above] at (4,2) { \begin{tabular}{c} Subgrids \\ \small (streamed) \end{tabular} };
    \node[above] at (2,2) { \begin{tabular}{c} Visibilities \\ \small (streamed) \end{tabular} };
    \foreach \i in {1,2,3,4}{
        \node[right] at (0, 5-\i-2.5) { Node \i};
        \node[rectangle, draw, minimum width=.9cm, minimum height=.9cm, inner sep=0] (facet\i) at (7.5,5-\i-2.5) {
            \tiny $\mathcal F[\B_\i\!\!\ast\!\!\V]$
        };
    }
    \node at (.5, -1.8) { \tiny\vdots };

    \foreach \i in {1,...,8}{
      \node[rectangle, draw, minimum height=.5cm] (grid\i) at (4,2.25-0.5*\i) {
        \tiny $\A_\i \V$
      };
        \node[rectangle, draw, minimum width=.9cm, minimum height=.2cm] (visout\i) at (2,2.35-0.5*\i) {};
        \draw[{Latex[length=1mm]}-,transform canvas={yshift=.075em}] (grid\i) -- (visout\i);
        \draw[-{Latex[length=1mm]},transform canvas={yshift=-.075em},dashed] (grid\i) -- (visout\i);
        \node[rectangle, draw, minimum width=.9cm, minimum height=.2cm] (visoutb\i) at (2,2.15-0.5*\i) {};
        \draw[{Latex[length=1mm]}-,transform canvas={yshift=.075em}] (grid\i) -- (visoutb\i);
        \draw[-{Latex[length=1mm]},transform canvas={yshift=-.075em},dashed] (grid\i) -- (visoutb\i);
    }
    \foreach \i in {1.5,2.5,3.5,4.5}{
        \draw[dotted] (0,\i-2.5) -- (8,\i-2.5);
    }
    \foreach \i in {1,2,3,4}{
        \foreach \j in {1,...,8}{
            \draw[{Latex[length=1mm]}-,transform canvas={yshift=.075em}] (facet\i) -- (grid\j);
            \draw[-{Latex[length=1mm]},dashed,transform canvas={yshift=-.075em}] (facet\i) -- (grid\j);
        }
    }
    \node[circle, draw, fill=white, minimum size=1cm, inner sep=-3] at (5.75,0)
         { \small\begin{tabular}{c} \tiny swiFTly \end{tabular} };
     \node[circle, draw, fill=white, minimum size=1cm, inner sep=0] at (3.125,0.5)
              { \tiny grid };
     \node[circle, draw, dashed, fill=white, minimum size=1cm, inner sep=0] at (3.125,-0.5)
              { \tiny degrid };
  \end{tikzpicture}

  \caption{Distribution concept sketch of our algorithm.  The visibilities are gridded to subgrids (solid lines),
      and the   contributions are accumulated to facets. The dashed lines indicate 
    the reverse direction:  the contributions are extracted from the facets,
      and   the visibilities are degridded from the subgrids.}
    \label{fig:theory}

\end{figure}

Therefore, this paper focuses on the Fourier transform specifically,
describing a scalable algorithm that addresses the core challenges of
large-scale interferometric processing. Our method fully distributes the image
data as well as the computation, never needing to assemble the entire image or
$uv(w)$-grid at any stage. As with previous work, this is achieved using a distributed
facet approach, so different parts of the image are held and processed in
memory separately.  However, our method solves the conceptually easier problem of
transforming them from and to cut-outs of the full-resolution spatial-frequency
grid, which we call `subgrids'. Hence, in our approach both the image-plane and
the aperture-plane are divided into smaller parts that can be processed in
parallel. This can be used to distribute conventional visibility (de)gridding algorithms
as a separate step, as shown in
\hyperref[fig:theory]{Fig. \ref*{fig:theory}}.  This would process every
visibility only once on a single node, with a roughly predictable processing
order.

All we need is an efficient way to transform between such subgrids and
facets, which would make it a form of distributed Fourier transform.
Distributed fast Fourier transform (FFT)
algorithms\,\citep[e.g.][]{frigo1998fftw} might seem the obvious choice;
however, they are not well suited for our purposes. For non-coplanarity
corrections we are often only interested in local regions of the $uvw$
spatial-frequency space, and  therefore we need to minimise not only the computational
work spent on unneeded $uvw$ regions, but also the time the full grid or
image data is held in memory. Additionally, we
would like to apply extra image-space factors per subgrid, for example for  
widefield corrections, which further pushes us towards a different class of
algorithm.

Our solution is therefore to stream portions of spatial-frequency space to
worker nodes (or, during de-gridding, from them) exactly as and when required
to cover the telescope baselines: an incremental semi-sparse Fourier
transform algorithm for contiguous partitions of image and frequency
space. In \autoref{sct:Fundamentals} we define the core algorithm, which we
then extend in \autoref{sct:noncoplanarity} to cover the effects of Fresnel
diffraction when observing widefield. In \autoref{sct:Implementation} we
demonstrate how to assemble and parametrise the algorithm, which allows us to
perform scaling and performance tests using SKA-scale parameters in
\autoref{sct:Results}. We finally wrap things up with a discussion in
\autoref{sct:final}.

\section{Core algorithm}
\label{sct:Fundamentals}

Interferometric imaging measures complex visibilities $V$,
and relates them to the planar projection of the sky intensity distribution $I$ as 
\begin{align}
V(u,v,w)
= \iint
I(l,m)\, e^{-2\pi iw(\sqrt{1- l^2- m^2}-1)}
e^{-2\pi i(ul+vm)}\,\text{d}l\,\text{d}m,
\end{align}

\noindent where $l$ and $m$ are sky direction cosines, and $u$, $v$,
and $w$ are coordinate components of telescope baselines\,\citep[compare e.g.][]{cornwell1992radio}.
The right-hand side involves a Fourier transform,
which is why an efficient approximation of $I$ from visibility samples
$V$ (or vice versa) requires discrete Fourier
transforms.

\subsection{Notation}
\label{sct:model}
\newcommand{\smallsum}{{\textstyle\sum}}
For simplicity we limit ourselves to one-dimensional functions for the moment.
By convention, all named functions are in
visibility space (i.e. using spatial-frequency $u$ coordinate).
The image or gridded visibilities to be transformed are represented
by a function $\V$, no matter the transformation direction considered.
Furthermore, all functions are discrete using a given step rate of
$2\yV$ and repeating with a period of $2\xV$ satisfying $4\xV\yV \in \mathbb Z$.
This allows us to define a `scaled'
discrete Fourier transform as
\begin{align}
\mathcal F[\mathrm f](l) &=
  \frac{1}{2\xV}
  \sum_{\substack{-\xV \le \xx < \xV\\2\xx\yV \in \mathbb Z}}
  \mathrm f(\xx) \,e^{-2\pi i \xx\yy}
  &&\text{ for } 2 l \xV \in \mathbb Z.
\end{align}
This makes $\mathcal F \mathrm f$ an
image-space function ($\yy$
coordinate) that is sampled with a step rate of $2\xV$ and repeats
with a period of $2\yV$.
We   use the following notations and properties:
\begin{align}
(\mathrm f \!+\! \mathrm g)(u) &= \mathrm f(u) + \mathrm g(u)
  &\Rightarrow&&
\mathrm f + \mathrm g &= \mathcal F^{-1}[\mathcal F\mathrm
  f+\mathcal F\!\mathrm g] ,\\
(\mathrm f \mathrm g)(u) &= \mathrm f(u) \mathrm g(u)
  &\Rightarrow&&
\mathrm f \mathrm g &= \tfrac{1}{2\yV} \mathcal F^{-1}[\mathcal F\mathrm f\!\ast\!\mathcal F\!\mathrm g] ,\\
(\mathrm f \!\ast\! \mathrm g)(u)  &= \!\!\!\!\sum_{\substack{-\xV \le \tau < \xV\\ 2\tau\yV \in \mathbb Z}}\!\!\!\!\mathrm f(\tau) \mathrm g(u\!-\!\tau)
&\Rightarrow&&
\mathrm f \!\ast\! \mathrm g &= 2\xV\mathcal F^{-1}[\mathcal F\mathrm f\ \mathcal F\!\mathrm g].
\end{align}

\noindent
For this paper we   often need to deal with functions
that are sampled more coarsely than the base $\xV$/$\yV$ discretisation.
We   use Kronecker comb functions to represent this:
\begin{align}
  \Sh_{2\xf}(u) = \sum_{k=-\infty}^{\infty} \delta(u - k 2\xf)
  \;\Rightarrow\;
  \mathcal F[\Sh_{2\xf}] = \tfrac{1}{2\xf} \smpy{2\xf}.
\end{align}
Here $2\xf\yV, 2\xf\xV^{-1} \in \mathbb Z$. Multiplication
with $\Sh_{2\xf}$ samples frequency space at a rate of $(2\xf)^{-1}$,
while convolution with $\Sh_{2\xf}$ samples image space at a rate of $2\xf$.
With $2\xV\yf, 2\yf\yV^{-1}, 4\xf\yf \in \mathbb Z$ we obtain
\begin{align}
 \mathrm f = \tfrac{\xV}{\xf}
  \smpx{2\xf} \ast \smpy{2\yf} \mathrm f \;\;\Leftrightarrow\;\;
  \mathcal F\mathrm f
  = \tfrac{\yf\xV}{\xf\yV}(\Sh_{2\yf} \ast \Sh_{(2\xf)^{\shortminus1}} \mathcal F \mathrm f);
  \label{eq:frepresent}
\end{align}
\noindent
in other words,  where a function $\mathrm f$ repeats with a period of $2\xf$ and
is non-zero at a rate of $2\yf$, its Fourier transform repeats with a
period of $2\yf$ and has a rate of $2\xf$. For
numerical purposes, \autoref{eq:frepresent} furthermore
shows that we can represent $\mathrm f$ or $\mathcal F \mathrm f$
entirely using exactly $|4\xf\yf|$ samples.

\subsection{Problem statement}
\label{sct:problem}

The distributed Fourier transform problem is now given by mask
functions $\A_i$ and $\B_j$, which encode how
we wish to partition $\V$ in spatial-frequency and image space:
\begin{align}
  \sum_i \A_i \V &= \V &&\text{ and }
  &\sum_j (\B_j \ast \V) &= \V
\end{align}
with $\xA < \xV$ and $\yB < \yV$ such that we can represent subgrids
$\A_i \V$ using $4\xA\yV$ samples (subgrid size) and facets $\B_j
\ast \V$ using $4\xV\yB$ samples (facet size).  We  expect
$\A_i$ to be boxcar filters in frequency space and $\B_j$ boxcar
filters in image space  (blue graphs in
\autoref{fig:overview}), and therefore with $\xV\xA^{-1},
\yV\yB^{-1}\in\mathbb{Z}$
\begin{align}
\A_i\V = \A_i(\smpx{2\xA}\!\ast\!\A_i\V), \,\,
\B_j\!\ast\!\V = \tfrac{\yV}{2\xV\yB} \B_j\ast\smpy{2\yB}(\B_j\!\ast\!\V)
\end{align}
by Whittaker–Shannon interpolation.
As illustrated in \hyperref[fig:theory]{Fig. \ref*{fig:theory}}, now the goal
is to reconstruct $\B_j \ast \V$ given all
$\A_i \V$. On paper this is straightforward due to
linearity:
\begin{align}
\B_j \ast \V = \B_j \ast \sum_i \A_i \V
= \sum_i \left(\B_j \ast \A_i \V \right).
\end{align}
So the contribution of subgrid $i$ to facet $j$ is simply
$\B_j \ast \A_i \V$.
For an efficient distributed algorithm, this contribution
would have to  be computed and exchanged between nodes.
Unfortunately, na\"ive attempts to represent
$\B_j\ast\A_i\V$ using $4\xA\yB$ samples would fail, as we cannot apply
Whittaker-Shannon simultaneously in image and spatial-frequency space:
\begin{align}
  \tfrac{\yV}{2\xV\yB}
  \B_j\ast\A_i\smpy{2\yB}(\smpx{2\xA}\ast\B_j\ast\A_i\V)
  \neq   \B_j\ast\A_i\V.
\end{align}
The reason is that $\B_j$ never approaches zero in frequency space (see
\autoref{fig:overview}, bottom left plot), and therefore neither does
$\B_j\ast\A_i\V$. Hence, tiling it with a period of $2\xA$
effectively loses information.

\begin{figure*}[t]
\begin{tikzpicture}

\definecolor{darkgray176}{RGB}{176,176,176}
\definecolor{darkorange25512714}{RGB}{255,127,14}
\definecolor{forestgreen4416044}{RGB}{44,160,44}
\definecolor{lightgray204}{RGB}{204,204,204}
\definecolor{steelblue31119180}{RGB}{31,119,180}

\begin{groupplot}[
    group style={
      group size=2 by 2,
      vertical sep=0.12cm,
      horizontal sep=0.2cm,
      group name=overview
    },
    title style={font=\tiny,yshift=-1.5ex},
    height=.45\columnwidth, width=7cm,
    legend cell align={left},
    legend style={font=\tiny, fill opacity=0.8, draw opacity=1, text opacity=1},
    ymode=log,
    tick align=outside,
    tick pos=left,
    xticklabels={,,},
    ytick={1e-7,1e-6,1e-5,1e-4,1e-3,1e-2,1e-1,1,1e1,1e2,1e3,1e4,1e5,1e6,1e7},
    yticklabels=\empty,
    minor ytick={
      5e-8, 5e-7, 5e-6, 5e-5, 5e-4, 5e-3, 5e-2, 5e-1, 5e0, 5e1, 5e2, 5e3, 5e4, 5e5, 5e6, 5e7
    },
    ymin=2e-8, ymax=9e7,
    xmajorgrids=true,
    yticklabel shift=-3pt,
  ]

\nextgroupplot[
  title={spatial frequency space},
  xmin=-0.5, xmax=0.5,
  xtick={-0.3125,-0.3,-0.1,-0.0875},
  yticklabels={,$10^{-6}$,,,,,,$1$,,,,,,$10^6$},
]
\path[thin,draw=gray] (axis cs:-0.5,1)--(axis cs:0.5,1);

\addplot [thick, steelblue31119180] table {tikz/overview-000.dat};
\addlegendentry{\tiny $|\A_i|$}
\addplot [semithick, darkorange25512714] table {tikz/overview-001.dat};
\addlegendentry{\tiny $|\m_i|$}

\nextgroupplot[
  title={image space},
  xmin=-1000, xmax=1000,
  xtick=\empty
]
\path[thin,draw=gray] (axis cs:-1000,1)--(axis cs:1000,1);
\addplot [thick, steelblue31119180] table {tikz/overview-002.dat};
\addlegendentry{\tiny $|\mathcal F\A_i|$}
\addplot [semithick, darkorange25512714] table {tikz/overview-003.dat};
\addlegendentry{\tiny $|\mathcal F\m_i|$}

\nextgroupplot[
  xmin=-0.5, xmax=0.5,
  xtick={-0.3125,-0.3,-0.1,-0.0875,-0.0125,0.0125},
  xmajorgrids=false,
  yticklabels={,$10^{-6}$,,,,,,$1$,,,,,,$10^6$},
]
\path[thin,draw=gray] (axis cs:-0.5,1)--(axis cs:0.5,1);

\path[thin,densely dotted,draw=gray!50!white] (axis cs:-0.3125,1e-8)--(axis cs:-0.3125,1e8);
\path[thin,densely dotted,draw=gray!50!white] (axis cs:-0.3,1e-8)--(axis cs:-0.3,1e8);
\path[thin,densely dotted,draw=gray!50!white] (axis cs:-0.1,1e-8)--(axis cs:-0.1,1e8);
\path[thin,densely dotted,draw=gray!50!white] (axis cs:-0.0875,1e-8)--(axis cs:-0.0875,1e8);
\path[thin,draw=gray!50!white] (axis cs:-0.0125,1e-8)--(axis cs:-0.0125,1e8);
\path[thin,draw=gray!50!white] (axis cs:0.0125,1e-8)--(axis cs:0.0125,1e8);

\coordinate (xml) at (axis cs: -0.3125,1e-8);
\coordinate (xal) at (axis cs: -0.3,1e-8);
\coordinate (xar) at (axis cs: -0.1,1e-8);
\coordinate (xmr) at (axis cs: -0.0875,1e-8);

\coordinate (xnl) at (axis cs: -0.0125,1e-8);
\coordinate (xnr) at (axis cs: 0.0125,1e-8);

\addplot [thick, steelblue31119180] table {tikz/overview-004.dat};
\addlegendentry{\tiny $|\B_j|$}
\addplot [semithick, darkorange25512714] table {tikz/overview-005.dat};
\addlegendentry{\tiny $|\n_j|$}
\addplot [semithick, forestgreen4416044] table {tikz/overview-006.dat};
\addlegendentry{\tiny $|\nb_j|$}

\nextgroupplot[
  xmin=-1000, xmax=1000,
  xtick={-240,-200,200,240}
]
\path[thin,draw=gray] (axis cs:-1000,1)--(axis cs:1000,1);

\coordinate (ynl) at (axis cs: -240,1e-8);
\coordinate (ybl) at (axis cs: -200,1e-8);
\coordinate (ybr) at (axis cs: 200,1e-8);
\coordinate (ynr) at (axis cs: 240,1e-8);

\addplot [thick, steelblue31119180] table {tikz/overview-007.dat};
\addlegendentry{\tiny $|\mathcal F\B_j|$}
\addplot [semithick, darkorange25512714] table {tikz/overview-008.dat};
\addlegendentry{\tiny $|\mathcal F\n_j|$}
\addplot [semithick, forestgreen4416044] table {tikz/overview-009.dat};
\addlegendentry{\tiny $|\mathcal F\nb_j|$}
\addplot [semithick, densely dotted, forestgreen4416044] table[y expr=1/\thisrow{y}] {tikz/overview-008.dat};
\addlegendentry{\tiny $|\mathcal F\n_j|^{-1}$}

\end{groupplot}

\node[left, align=left, text=black] at (-.6,0) {
  \fbox{\fbox{\begin{tabular}{cl}
    $\V$ & Image / visibility data \emph{(input)} \\[4pt]
    $\A_i$ & Subgrid masks \emph{(input)} \\
    $\m_i$ & Padded subgrid masks \\
    & \emph{(such that $(1-\m_i)(\n_j\ast\A_i) \approx 0$)} \\[4pt]
    $\B_j$ & Facet masks \emph{(input)} \\
    $\n_j$ & Smoothed facet mask \emph{(chosen)}  \\
    $\nb_j$ & Inverse of $\n_j$ wrt $\B_j$ \\
    & \emph{(such that $\n_j \ast \nb_j = \B_j$)} \\[2pt]
    \multicolumn{2}{c}{$\Downarrow$} \\[2pt]
    \multicolumn{2}{c}{$\begin{aligned}
      \nb_j\ast\m_i(\n_j \ast \A_i \V) &\approx \B_j\ast\A_i\V \\
      \A_i(\n_j\ast\m_i(\nb_j \ast \V)) &\approx \A_i(\B_j\ast\V)
      \end{aligned}$
    }
  \end{tabular}}}
};

\draw[<->]
  ($(ynl)-(0,11pt)$) --
  node (ynlbl) [fill=white, inner sep=1pt] {\tiny \phantom{$2\yn$}}
  ($(ynr)-(0,11pt)$);
\draw[<->]
  ($(ybl)-(0,4pt)$) --
  node[fill=white, inner sep=1pt] {\tiny $2\yB$}
  ($(ybr)-(0,4pt)$);
\node at (ynlbl) {\tiny $2\yn$};

\draw[<->]
  ($(xnl)-(0,4pt)$) --
  node[fill=white, inner sep=1pt,below=2pt] {\tiny $2\xn$}
  ($(xnr)-(0,4pt)$);

\draw[<->]
  ($(xml)-(0,11pt)$) --
  node (ynlbl) [fill=white, inner sep=1pt] {\tiny \phantom{$2\xm$}}
  ($(xmr)-(0,11pt)$);
\draw[<->]
  ($(xal)-(0,4pt)$) --
  node[fill=white, inner sep=1pt] {\tiny $2\xA$}
  ($(xar)-(0,4pt)$);
\node at (ynlbl) {\tiny $2\xm$};
  
\end{tikzpicture}%
    \caption{Key functions and relationships at a glance}
  \label{fig:overview}
\end{figure*}

\subsection{Approximation}
\label{sct:approximation}
\label{sct:errors}

\begin{figure}[b]
    \centering

\definecolor{color0}{rgb}{0.12156862745098,0.466666666666667,0.705882352941177}
\definecolor{color1}{rgb}{1,0.498039215686275,0.0549019607843137}
\definecolor{color2}{rgb}{0.172549019607843,0.627450980392157,0.172549019607843}

\begin{tikzpicture}
\pgfplotsset{
  every axis legend/.append style={
    at={(0.5,1.03)},
    anchor=south
  }
}

\begin{groupplot}[
    group style={group size=1 by 1,vertical sep=0.15cm,
                 xticklabels at=edge bottom},
    legend style={font=\tiny},
    height=.45\columnwidth,
    tick align=outside,
    xmin=-120, xmax=120,
    width=\columnwidth,
    no markers,
    xtick={-60,-50,50,60},
    xticklabels={
      \tiny$\pdfliteral{ 1 0  0 1 0 -1 cm}\!-\yn$,
      \tiny$\pdfliteral{ 1 0  0 1 0 1 cm}\;-\yB$,
      \tiny$\pdfliteral{ 1 0  0 1 0 1 cm}\yB$,
      \tiny$\pdfliteral{ 1 0  0 1 0 -1 cm}\yn$},
]

\nextgroupplot[ymode=log,ymin=2e-10, legend columns=2, legend style={draw=none},ytickten={-3,-5,-7,-9}]

\pgfplotstableread{tikz/facet-subgrid-image.csv}\loadedtable
\pgfplotstableread{tikz/facet-subgrid-image-worst.csv}\loadedtableworst

\addplot+[semithick,color=color0] table[x={l},y={error1}] {\loadedtable};
\addplot+[semithick,color=color0!50!white] table[x={l},y={error1}] {\loadedtableworst};
\addplot+[semithick,dotted,domain=-256:256] {1.5e-4/abs(x+60)};
\addplot+[semithick,color=color1] table[x={l},y={error2}] {\loadedtable};
\addplot+[semithick,color=color1!50!white] table[x={l},y={error2}] {\loadedtableworst};

\path[draw, dashed] (axis cs:60,1.5e-10)--(axis cs:60,5e-2);
\path[draw, dashed] (axis cs:50,1.5e-10)--(axis cs:50,5e-2);
\path[draw, dashed] (axis cs:-60,1.5e-10)--(axis cs:-60,5e-2);
\path[draw, dashed] (axis cs:-50,1.5e-10)--(axis cs:-50,5e-2);


\legend{
$\big|\mathcal F\big[(1\!-\!\mathrm{m}_i)(\mathrm{n}_j\!\ast\!\mathrm{A}_i \V)\big]\big|\quad$,,,
$\tiny\big|\mathcal F\big[\mathrm{b}_j\!\ast\!(1\!-\!\mathrm{m}_i)(\mathrm{n}_j\!\ast\!\mathrm{A}_i \V)\big]\big|$}

\end{groupplot}

\end{tikzpicture}%
    \caption{Image space error (light colours: worst case $\V$)}
    \label{fig:fbag}
\end{figure}

The key idea is to use a different representation. We suppose a function
$\n_j$ that falls close to zero in both frequency
and image space (with associated limits $\xn$ and $\yn$
respectively). Then the same should be true for $\n_j \ast \A_i\V$,
so we should be able to say
\begin{align}
\m_i(\n_j\ast\A_i\V) \approx \n_j\ast\A_i\V,
\end{align}
where $\m_i$ is a boxcar filter with $\xm > \xA + \xn$ (\autoref{fig:overview}, left plots). If we
furthermore find an `inverse' function $\nb_j$ that gets us back to
$\B_j=\nb_j\ast\n_j$ (which will require $\yn \ge \yB$), it follows that
\begin{align}
    \B_j\ast\V = \nb_j\ast\sum_i(\n_j\ast \A_i\V)
\approx \nb_j\ast\sum_i\m_i(\n_j\ast \A_i\V).
\end{align}

\noindent
This approach mirrors visibility gridding. If $\A_i$ were delta
functions identifying positions of visibilities (oversampled at a
rate of $2\yV$), then $\n_j$ would be the gridding kernel, $\m_i$ the
grid environment around each visibility, and $\nb_j$ the gridding
correction. What we are doing is simply `bulk re-gridding' to
coarser facet grids.

This suggests that window functions like prolate spheroidal wave
functions (PSWFs; see  \autoref{fig:overview}) are a good choice
for $\n_j$.
Unfortunately, there is no window function that is perfectly limited
in both image and frequency space (uncertainty principle), so there will
always be errors. We can quantify the effects by subtracting both
sides from above for a subgrid $i$:
\begin{align}
  \nb_j \ast (1 - \m_i) (\n_j \ast \A_i \V) \approx 0.
\end{align}
\autoref{fig:fbag} shows what these errors look like in image space
for a centred facet.  Before convolution with $\nb_j$ (dark blue
line) the absolute error is mostly constant with small peaks at $\pm
l_\n$. As the light blue line shows, this only becomes slightly worse
for a worst-case $\V$ (single source at $-l_\n$).
On the other hand, convolution with $\nb_j$ introduces a distinct
U-shaped error pattern. The reason is that for $\n_j$ to be
approximately limited in spatial-frequency space, it needs to be
smooth in image space,  and as we require $\mathcal F\n_j$ to fall to
zero, the inverse $|\mathcal F\n_j|^{-1}$ tends to infinity at
$\pm\yn$ (see also \autoref{fig:overview}). This is why we need to
somewhat over-dimension $\n_j$ in image-space ($\yn > \yB$) to keep
error magnification in check.

\subsection{Method}
\label{sct:proof}

At this point we know that $\n_j\ast \A_i\V$ is approximately limited to an
$\xm$ region in frequency space, so assuming $\xV\xm^{-1},
4\xm\yn \in\mathbb Z$:
\begin{align}
  \nb_j \ast \tfrac{\yV}{\yn}\smpy{2\yn} (\n_j\ast \A_i\V) &= \B_j\ast \A_i\V \\
\Rightarrow
  \nb_j \ast \m_i\, \tfrac{\yV}{\yn}\fbox{$\smpy{2\yn} (\smpx{2\xm} \ast \n_j\ast \A_i\V)$} &\approx \B_j\ast \A_i\V
  \label{eq:contribution}
\end{align}
by associativity of multiplication. Therefore $\n_j\ast\A_i\V$ can
be represented approximately using $4\xm\yn$ samples (boxed
expression), solving the core challenge from \autoref{sct:problem}.

If we define $\n_j^0$ as a mask representing the
image-space extent of $\n_j$ ($\mathcal F \n_j^0(\yy) = 1$ exactly
where $\mathcal F\n_j(\yy) \ne 0$), we can also show how to compute this
contribution efficiently,
\begin{align}
  &\smpy{2\yn} \left(\smpx{2\xm} \ast \n_j\ast \A_i\V\right) \nonumber \\
  &= \tfrac{\xm\yV}{2\xV^2\yn} \smpy{2\yn} (\smpx{2\xm} \ast \n_j) \ast \smpy{2\yn} (\n_j^0 \ast \smpx{2\xm} \ast \A_i\V),
\end{align}
as $\smpx{2\xm}=\frac{\xm}{\xV}\smpx{2\xm}\ast\smpx{2\xm}$,
 $\n_j = \frac1{2\xV} \n_j \ast \n_j^0$, and $\smpy{2\yn} (\mathrm f \ast \mathrm g) =
  \frac{\yV}{\yn} \smpy{2\yn} \mathrm f \ast \smpy{2\yn} \mathrm g$, where
  $\mathrm f$ and $\mathrm g$ are zero in image space outside a $2\yn$
  region.  This means that if we have padded subgrid data $\smpx{2\xm} \ast
\A_i\V$ ($4\xm\yV$ samples; $\xm > \xA$) in image space, then we just need to
select $4\xm\yn$ samples where we know $\mathcal F\n_j(l) \ne 0$, multiply by
$\mathcal F\n_j$ (sampled at the same points), and we have calculated our
subgrid contribution representation.

Applying the contribution is equally straightforward:
\begin{align}
  &\smpy{2\yn} \sum_i (\B_j\ast \A_i \V) {\nonumber} \\
  &\approx \smpy{2\yn} \sum_i \nb_j \ast \m_i\, \tfrac{\yV}{\yn}\smpy{2\yn} (\smpx{2\xm} \ast \n_j\ast \A_i\V) \\
  &=  \tfrac{\yV}{\yn}\smpy{2\yn} \nb_j\!\ast\!\sum_i (\smpy{2\yn} \m_i)\; \fbox{$\smpy{2\yn} ( \smpx{2\xm}\!\ast\!\n_j\!\ast\!\A_i\V)$}
\end{align}
substituting in the term from \autoref{eq:contribution}, then using
  $\smpy{2\yn} = \smpy{2\yn}\smpy{2\yn}$ and
  $\smpy{2\yn} (\mathrm f \ast \smpy{2\yn} \mathrm g) =
  \smpy{2\yn} \mathrm f \ast \smpy{2\yn} \mathrm g
  $, where $\mathrm f$ is zero in image space outside a $2\yn$
  region. This means that we
need to pad in frequency space from $4\xm\yn$ to $4\xm\yV$ samples, then
multiply the sum in image space by an equivalently sampled $\nb_j$.

\subsection{Shifts}
\label{sct:shifts}
\label{sct:shifts2}

\begin{figure*}[t]
  \centering
  \input{tikz/2d_illustration.tex}%
  \caption{2D algorithm sketch of both transformation directions, highlighting
    the symmetry. This also illustrates
    buffer sizes, showing how padding facet data to $4\xV\yn$ tends to
    dominate memory consumption both in the horizontal (u) and vertical (v) axes.
    This is why it is important to share intermediate buffers for subgrids
    from the same `column' ($\m\ui = \m^\text{(u)}_{i'}$).
    }
  \label{fig:algorithm2D}
\end{figure*}

For simplicity let us assume that
all $\m_i$ and $\n_j$ are the same functions, just shifted by
subgrid or facet offsets $\dxi$ / $\dyj$ as follows:
\begin{align}
\m_i &= \sii \ast \m ,\\
\n_j &= (\tji) \n \quad = \mathcal F^{\shortminus 1}[\delta_{\dyj}\ast\mathcal F\n].
\end{align}

\noindent As $\sii\ast \mathrm{f}\hspace{.01cm}\mathrm{g} = (\sii
\ast \mathrm f)(\sii \ast \mathrm{g})$, $\tji
(\mathrm{f}\ast\mathrm{g}) = \tji\mathrm{f}\,\ast\,\tji\mathrm{g}$,
and $\si \ast
   \tj = \tj$ (assuming $\dxi\dyj \in \mathbb Z$),
we can derive
\begin{align}
\;
&\tj (\B_j \ast \V)  ,\\ &\approx
 \tj \nb_j \ast \sum_i \left( \sii \ast \m
 \left(\n \ast (\si \ast \tj) (\si \ast \A_i \V) \right) \right) ,\\
 &=
 \tj \nb_j \ast \sum_i \left( \sii \ast \m
 \left(\n \ast \tj (\si \ast \A_i \V) \right) \right).
\end{align}

\noindent
This tells us that we only need to have
one sampled representation of $\m$ and $\n$ because  the function offsets applied to the data
  simply correspond to index shifts for
image or frequency space samples
as long as $2\xm\dyj, 2\dxi\yn \in \mathbb Z$.
We   revisit these side conditions in
\autoref{sct:dim_shifts}.

\subsection{Two dimensions}
\label{sct:2D}

We can easily generalise our reasoning to images simply by
re-defining $\V,\A,\B,\m,\nb$, and $\n$ to be two-dimensional
functions. We additionally
assume $\m_i$ and $\nb_j$ to be separable,  
\begin{align}
\m_i = \m\ui \m\vi, \quad
\nb_j = \nb\uj \ast \nb\vj,
\end{align}
such that $\m^\text{(u)}$ and $\m^\text{(v)}$ are constant along the v- and u-axes, respectively, and $\nb^\text{(u)}(u,v)=0$ for $v \ne 0$ and
$\nb^\text{(v)}(u,v)=0$ for $u \ne 0$.  Then we can re-order as follows:
\begin{align}
\B_j \ast\V \approx
  \nb\uj \ast \sum_{i} \m\ui \left( \nb\vj \ast \m\vi
    \big(\n_j\ast \A_i \V\big)
  \right).
\end{align}
\vspace{-.3cm}

\noindent
Convolution with $\nb\vj$ straight after multiplication with $\m\vi$
is a small optimisation. An implementation can discard $4\xV(\yn-\yB)$ rows where
$\mathcal F\nb\vj$ is zero by doing the Fourier transform $\mathcal
F^{(v)}$ along the $v$-axis right away. The sum can be pulled further inwards if
we assume that some subgrids are in the same column (i.e. share a
certain value of $\m\ui$). Then with $M\uu$ the set of all $\m\ui$:
\begin{align}
\B_j \ast\V \approx
\nb\uj \ast \sum_{\m\uu \in M\uu} \m\uu \Big( \nb\vj \ast
  \sum_{i,\,\m\ui = \m\uu_{\phantom{i}}} \m\vi
    \big(\n_j\ast \A_i \V\big)
  \Big).
\end{align}
As illustrated in \hyperref[fig:algorithm2D]{Fig. \ref*{fig:algorithm2D}a}, this means we can accumulate contributions from
entire columns of subgrids, reducing the number of Fourier transforms 
needed.

\subsection{Dual variant}
\label{sct:duality}

As degridding relates to gridding, there is a similar dual
algorithm variant for going from facets to subgrids. Analogously to the steps taken in 
\autoref{sct:model}, we start with the observation that
\begin{align}
\A_i \V = \A_i \,\Big(\sum_j \B_j \ast \V\Big)
= \sum_j \A_i \left(\B_j \ast \V \right),
\end{align}

\noindent
and we then decompose $\B_j = \n_j\ast \nb_j$ and
$\A_i = \A_i \m_i$ as done in
\autoref{sct:approximation} to approximate the sum term
\begin{align}
\A_i \left(\B_j \ast \V \right)
= \A_i \m_i \left(\n_j \ast \nb_j \ast \V \right)
\approx \A_i \left(\n_j \ast \m_i (\nb_j \ast \V ) \right)
\end{align}
with the error term $\A_i \big(\n_j \ast (1 - \m_i) (\nb_j \ast \V )\big)$,
which behaves equivalently to the errors discussed in
\autoref{sct:errors}.  This also works in 2D:
\begin{align}
\A_i \V \approx
\A_i \sum_j
\left( \n_j \ast \m\vi
\left( \nb\vj \ast \m\ui
\left( \nb\uj \ast \V \right)\right)\right).
\end{align}

\noindent
The optimisation from the previous   section also has an equivalent here:
\begin{align}
\A_i \V \approx \A_i \sum_j
\left( \n_j \ast \mathrm{T}_{j,\m\ui}\right), \quad
\mathrm{T}_{j,\m\uu} =
\m\vi
\left( \nb\vj \ast \m\uu
\left( \nb\uj \ast \V \right)\right).
\end{align}
So we only need to calculate $\mathrm{T}_{j,\m\uu}$ once for every
combination of facet and subgrid column,
as shown in \hyperref[fig:algorithm2D]{Fig. \ref*{fig:algorithm2D}b}.

\section{Handling widefield effects}
\label{sct:noncoplanarity}

A general-purpose discrete Fourier transform
algorithm cannot quite cover the full complexity of the measurement equation from
\autoref{sct:Fundamentals} as wide fields of view require correction for telescope non-coplanarity ($w\ne0$) / sky curvature ($\n\ne1$).
As noted by \cite{cornwell2008noncoplanar}, we can represent this
using a convolution with a $w$-specific Fresnel diffraction pattern:
\begin{align}
  \g_w = \mathcal F_{l,m}^{-1}[ e^{2\pi w (\sqrt{1-l^2-m^2}-1)} ].
  \label{eq:gw}
\end{align}

\noindent
This convolution effectively introduces a third dimension into the problem, but
only on the spatial frequency side.  To represent this, we modify our subgrid
definition such that each subgrid $i$ has an arbitrary $w$-offset $\dwi$
associated with it:
\begin{align}
A_i (\g_{\dwi} \ast \V).
\end{align}
In this section we consider the problem variant of generating such $w$-shifted
subgrids from facets, or vice versa.  We note that even with $w$-offsets,
satisfying $\sum_i \A_i (\g_{\dwi} \ast \V) = \V$ (compare
\autoref{sct:problem}) is still possible by weighting $\A_i$. This is well
understood in radio astronomy, but is a distraction for the purpose of this
paper, so we stick to the dual algorithm (facet to subgrid; see
\autoref{sct:duality}) for this section.

\subsection{Implementation options}
\label{sct:wstacking}

If we simply replace $\V$ by $(\g_{\dwi}\!\ast\!\V)$ in our subgrid formula
from \autoref{sct:duality} we arrive at
\begin{align}
    \A_i (\g_{\dwi} \ast \V)
    \approx
    \A_i \sum_j \left(\n_j \ast \m_i (
    \nb_j \ast \g_{\dwi} \ast \V
    ) \right).
\end{align}
So the obvious way to obtain $w$-shifted subgrids with our algorithm is to
multiply facets by $\mathcal F\g_{\dwi}$, which is known as $w$-stacking
\citep[e.g.][]{offringa2014wsclean}.  This is simple and precise no matter the
field of view or value of $|\dwi|$.

On the other hand, this forces us to repeat the entire data re-distribution for
every needed value of $\dwi$. Sharing intermediate subgrid column buffers, as
shown in \hyperref[fig:algorithm2D]{Fig. \ref*{fig:algorithm2D}}, would now
especially require $\hat \xx_i$ and $\dwi$ to match.  We can still minimise the
amount of subgrid data to reproduce for any given $\dwi$ level by skipping
generation of unneeded subgrids, yet this is still not particularly efficient.

\label{sct:noncoplanar_late_convolve}

Fortunately we have additional options.
After all, we can  treat $\B'_{j,\dwi}
= \B_j \ast \g_{\dwi}$ like a modified facet mask, which means that we
can decompose, as in \autoref{sct:approximation}, into
 $\B'_{j,\dwi} = \n'_{j,\dwi} \ast \nb_j$ using
a modified window function $\n'_{j,\dwi} = \n_j \ast \g_{\dwi}$.
This yields
\begin{align}
\A_i \sum_j \left((\B_j \ast \g_{\dwi}) \ast \V)\right)
\approx
\A_i \sum_j \left((\n_j \ast \g_{\dwi}) \ast \m_i (
  \nb_j \ast \V
) \right).
\end{align}
This is attractive because $\m_i (\nb_j \ast \V)$ is the
distributed contribution term, so this introduces
$w$-shifts after data re-distribution and all image-size FFTs. However, this
can only work as long as
$\n'_{j,\dwi}$  still acts
like a window function for our purposes:
\begin{align}
\A_i \left((\n_j \ast \g_{\dwi}) \ast (1 - \m_i) \right)\approx 0.
\end{align}
So for this approach to hold up, we must now identify masks $\m_i$
compatible not only with $\A_i$ and all $\n_j$, but also given
$\dwi$.

\begin{figure}
    \centering
\begin{tikzpicture}

\definecolor{color0}{rgb}{0.12156862745098,0.466666666666667,0.705882352941177}
\definecolor{color1}{rgb}{1,0.498039215686275,0.0549019607843137}
\definecolor{color2}{rgb}{0.172549019607843,0.627450980392157,0.172549019607843}
\definecolor{color3}{rgb}{0.83921568627451,0.152941176470588,0.156862745098039}
\definecolor{color4}{rgb}{0.580392156862745,0.403921568627451,0.741176470588235}

\pgfplotsset{
compat=1.11,
legend image code/.code={
\draw[mark repeat=2,mark phase=2]
plot coordinates {
(0cm,0.02cm)
(0.15cm,0.02cm)        
(0.3cm,0.02cm)         
};%
}
}

\begin{axis}[
    height=0.35\columnwidth,
    legend cell align={left},
    legend style={ font=\tiny, draw=none, fill=none, at={(0.5,1.03)},
      /tikz/every even column/.append style={column sep=0.5cm},
      anchor=south, legend columns=-1},
    log basis y={10},
    tick align=outside,
    tick pos=left,
    width=.9\columnwidth,
    x grid style={white!69.0196078431373!black},
    xmin=-120, xmax=120,
    xtick style={color=black},
    y grid style={white!69.0196078431373!black},
    xtick={
      -30,
      -60.3022689155527,
      -96.198043021569,
      30,
      60.3022689155527,
      96.198043021569
    },
    xticklabels={
      \tiny $\pdfliteral{ 1 0  0 1 0 -2 cm}-\xn$,
      \tiny $\pdfliteral{ 1 0  0 1 0 -2 cm}-\xg$,
      \tiny $\pdfliteral{ 1 0  0 1 0 -2 cm}-\xgn$,
      \tiny $\pdfliteral{ 1 0  0 1 0 -2 cm}\xn$,
      \tiny $\pdfliteral{ 1 0  0 1 0 -2 cm}\xg$,
      \tiny $\pdfliteral{ 1 0  0 1 0 -2 cm}\xgn$
    },
    ymin=1.79490527575195e-11, ymax=1,
    ymode=log,
    ytick style={color=black, font=\tiny}
  ]
\addplot [semithick, color0]
table {tikz/noncoplanarity-000.tsv};
\addlegendentry{$\n$}
\addplot [semithick, color2]
table {tikz/noncoplanarity-002.tsv};
\addlegendentry{$\g_w\ast\n^0$}
\addplot [semithick, color3]
table {tikz/noncoplanarity-003.tsv};
\addlegendentry{$\g_w\ast\n$}
\addplot [black, dotted, forget plot]
table {tikz/noncoplanarity-005.tsv};
\node[fill=white,inner sep=0,text=gray] at (axis cs:0,3e-8)
     {\tiny base error };

\end{axis}
\end{tikzpicture}%
    \caption{Window interaction with non-coplanarity term}
    \label{fig:noncoplanarity}
\end{figure}

\begin{figure}
    \centering
\begin{tikzpicture}

\definecolor{white}{rgb}{1.0,1.0,1.0};
\definecolor{color0}{rgb}{0.12156862745098,0.466666666666667,0.705882352941177}
\definecolor{color1}{rgb}{1,0.498039215686275,0.0549019607843137}
\definecolor{color2}{rgb}{0.172549019607843,0.627450980392157,0.172549019607843}
\definecolor{color3}{rgb}{0.83921568627451,0.152941176470588,0.156862745098039}
\definecolor{color4}{rgb}{0.580392156862745,0.403921568627451,0.741176470588235}
\definecolor{color5}{rgb}{0.549019607843137,0.337254901960784,0.294117647058824}
\definecolor{color6}{rgb}{0.890196078431372,0.466666666666667,0.76078431372549}

\pgfplotsset{
compat=1.11,
legend image code/.code={
\draw[mark repeat=2,mark phase=2]
plot coordinates {
(0cm,0.02cm)
(0.15cm,0.02cm)        
(0.3cm,0.02cm)         
};%
}
}

\begin{axis}[
    height=.33\columnwidth,
    legend cell align={left},
    legend style={
      font=\tiny, draw=none, fill=none, at={(0.45,1.03)},
      anchor=south, legend columns=-1
    },
    tick align=outside,
    tick pos=left,
    width=.9\columnwidth,
    x grid style={white!69.0196078431373!black},
    xlabel={\tiny\(\displaystyle 4\xg\yn
      \)},
    xmin=0, xmax=160,
    xtick style={color=black, font=\tiny},
    y grid style={white!69.0196078431373!black},
    ylabel={\tiny\(\displaystyle 4(\xgn-\xg)\yn\)},
    ytick style={color=black}
  ]

\addplot [white] coordinates { (0, 0) };
\addlegendentry{$W=\tiny 4\xn\yn=$}
\addplot [semithick, color0] table {tikz/xgwn_trends-002.dat};
\addlegendentry{$\tiny 8.0$}
\addplot [semithick, color1] table {tikz/xgwn_trends-003.dat};
\addlegendentry{$\tiny 10.0$}
\addplot [semithick, color2] table {tikz/xgwn_trends-004.dat};
\addlegendentry{$\tiny 12.0$}
\addplot [semithick, color3] table {tikz/xgwn_trends-005.dat};
\addlegendentry{$\tiny 14.0$}
\addplot [semithick, color4] table {tikz/xgwn_trends-006.dat};
\addlegendentry{$\tiny 16.0$}
\addplot [semithick, color0, densely dotted] table {tikz/xgwn_trends-011.dat};
\addplot [semithick, color1, densely dotted] table {tikz/xgwn_trends-012.dat};
\addplot [semithick, color2, densely dotted] table {tikz/xgwn_trends-013.dat};
\addplot [semithick, color3, densely dotted] table {tikz/xgwn_trends-014.dat};
\addplot [semithick, color4, densely dotted] table {tikz/xgwn_trends-015.dat};
\end{axis}

\end{tikzpicture}%
    \caption{Non-coplanarity margin relative to highest chirp
      frequency for $\yy_\text{max}=0.2$ (solid:
      facet size $\yn=0.1$, dotted: $\yn=0.2$) }
    \label{fig:xgwn2_trends}
\end{figure}

\subsection{Margins}
\label{sct:wt-error}

Fortunately, it is not unreasonable to expect $\n_j \ast \g_{\dwi}$ to
be limited in frequency space.  Using the definition of
$\g_w$ from \autoref{sct:noncoplanarity}, we can determine the
frequency by derivation of the exponent:
\begin{align}
\frac{\delta}{\delta l} w\sqrt{1-l^2} = \frac{- w l}{\sqrt{1-l^2}}.
\end{align}

\noindent
Now we can use the fact that $\n_j$ is an image-space filter. If
$\yy_\text{max}$ is the maximum distance from the phase centre in the
facet $\n_j$, then the effective size of $\g_w$ for the purpose of
$\n_j \ast \g_w$ should be about $\xg = |w| \yy_\text{max}
(1-\yy_\text{max}^2)^{-\frac12}$ in frequency space.  As
\autoref{fig:noncoplanarity} shows, $\g_w \ast \n^0$ (with $\n^0$
again the mask version of $\n$) indeed falls off, though it is hard to
spot due to $\n^0$ causing sinc-like ripples (green
line). Fortunately, convolution with $\n$ suppresses this by
design, although while adding signal spread of its own
(red line).

As a result, if we define $\xgn$ as the point where $\n \ast
\g_{\dwi}$ reaches the base error level associated with $\n$, we
would generally expect $\xgn - \xg \approx\xn$. As
\autoref{fig:xgwn2_trends} shows, things are not quite that simple:
$\xgn$ first grows a bit more than expected, then eventually even
falls below $\xn$ for some parameters. These interactions are due to
$\n_j$ not just limiting, but also weighting different frequencies of
the chirp differently.  In practice, this means that  we  have to numerically or experimentally
determine the maximum acceptable $w$ from $\xA$ and $\xm$.

\subsection{\textit{w}-towers}
\label{sct:wtowers}

The $w$-distance limit is not a problem in practice, as due
to $\g_{w_1+w_2} = \g_{w_1}\ast\g_{w_2}$ we
can use $w$-stacking to reach a sufficiently close $\twi$ first:
\begin{align}
\A_i (\g_{\dwi} \ast \V)
\approx
\A_i \Big(\g_{\dwi - \twi} \ast \sum_j \left(\n_j \ast \m_i (
  \nb_j \ast \g_{\twi} \ast \V
) \right)\Big).
\end{align}
This combination is much more efficient than pure $w$-stacking, as we now only have to repeat
data re-distribution for every unique value of $\twi$. Furthermore, in 
summing up the $\n_j\ast \m_i(\nb_j\ast$$\g_{\twi}\ast$$\V)$
contributions for some $w$-stacking plane $\twi$,
we naturally end up in image space (see
last step in \hyperref[fig:algorithm2D]{Fig. \ref*{fig:algorithm2D}} and the right side of
\autoref{fig:wtowers}). Therefore, image-space
multiplication with $\mathcal F $$\g_{\dwi - \twi}$ is basically free. In the special case that we are interested in subgrids that are
equidistant in $w$ (and have matching $\m_i$), we can  iteratively build a `$w$-tower'
of subgrids using only one constant $\mathcal F \g_{\Delta w}$, as
illustrated in \autoref{fig:wtowers}. We still need one Fourier
transform for every `storey' of the tower; however, these are quite a
bit smaller than the image size, and can thus   be computed relatively
efficiently.

For subgrids at maximum height $|\dwi-\twi|$, only the central $\A_i$ region
(of maximum size $\xA$) will be accurate. This is subject to $\xm > \xgn+\xA$
as established in the last section, effectively sacrificing $4(\xgn-\xn)^2$
subgrid area to deal with non-coplanarity.  Thus, covering larger $w$-ranges
either means reducing the usable $uv$-area per subgrid $\xA$\footnote{The value
of $\xgn$ depends on $|\dwi-\twi|$, so the usable $uvw$-volume actually has the
shape of a bipyramid. This is not very useful for filling $uvw$-volumes though,
so we assume $w$-towers to cover `boxes'.} or introducing more $w$-stacking planes
($\twi$ values), and therefore full data re-distributions. We show in
\autoref{sct:Parameters} how we can handle this particular trade-off.

\subsection{Optimisation of \textit{w}-distribution}
\label{sct:w-snapshots}

The number of subgrids or $w$-tower storeys to generate is still a major cost
factor, so optimising the visibility $uvw$ distribution for a minimum $w$-range
is highly desirable. For instance, we can use a shear transformation as
proposed in \cite{cornwell2012wide} to transform visibility $(u,v,w)$
coordinates to $(u,v,w')$ as follows:
\begin{align}
  w' &= w - h_u u - h_v v ,\\
  l' &= l + h_u n = l + h_u (\sqrt{1-l^2-m^2}-1) ,\\
  m' &= m + h_v n = m + h_v (\sqrt{1-l^2-m^2}-1).
       \label{eq:snapshotcoord}
\end{align}
\noindent This allows us to choose $h_u$ and $h_v$ to minimise $|w'|$, while
solving the same problem due to $ul+vm+wn = ul'+vm'+w'n$.
We still have $n = \sqrt{1-l^2-m^2} - 1$, and therefore the definition
  of $\g_w$ from \autoref{sct:noncoplanarity} remains unchanged, except that we now
  sample in $l'$ and $m'$ for the purpose of $w$-stacking or $w$-towers.
  Using computer algebra we can solve for the original $l$ and $m$:
\begin{align}
  l &= \frac{l' (1+h_\text{v}^2) + h_\text{u} (1 - m' h_\text{v} - \sqrt{c})}{h_\text{u}^2+h_\text{v}^2+1},\\
  m &= \frac{m' (1+h_\text{u}^2) + h_\text{v} (1 - l' h_\text{u} - \sqrt{c})}{h_\text{u}^2+h_\text{v}^2+1},
\end{align}
\begin{align*}
  \text{where } c &= 1-2 l' h_\text{u}-2 m' h_\text{v} +2 l' h_\text{u} m' h_\text{v} \\
  & \qquad - l'^2 \big(h_\text{v}^2+1\big)-m'^2 \big(h_\text{u}^2+1\big).
\end{align*}
To first order $l\approx l'$ and $m \approx m'$, which means that this
transformation will change neither $\yy_\text{max}$ nor the required
subgrid margin $\xgn$ from \mbox{\autoref{sct:wt-error}} significantly until
quite close to the horizon, so there is little efficiency loss
associated with this optimisation. This in contrast to the variant described
in \mbox{\cite{offringa2014wsclean}}, where flat offsets in $l$ and $m$
directly increase both $\yy_\text{max}$ and $\xgn$, reducing  the maximum height of $w$-towers
and therefore efficiency much more rapidly.

\subsection{Convolutional (de)gridding}
\label{sct:gridding}
\begin{figure}
  \centering
  \begin{tikzpicture}[inner sep=1]
    \begin{axis}[
        xmin=-128,xmax=384,ymin=-128,ymax=384,width=7.5cm,height=8.5cm, clip=false,
        xlabel=u, ylabel=v, zlabel=w,
        xtick=\empty,ytick=\empty,ztick=\empty,
        axis line style={gray, dashed}, every axis label/.append style ={gray},
        yticklabel pos=right,
      ]

      \foreach \i in {0,100,...,500} {
        \addplot3[patch,patch type=rectangle,color=black,fill=white]
          coordinates { (0,0,\i) (128,0,\i) (128,128,\i) (0,128,\i) };
      }

      \draw[densely dotted] (16,112,0) -- (16,112,500);
      \foreach \i in {0,100,...,500} {
        \addplot3[patch,patch type=rectangle,color=black,thick, fill=gray!50!white,
                  semitransparent]
          coordinates { (16,16,\i) (112,16,\i) (112,112,\i) (16,112,\i) };
      }

      \draw[densely dotted] (16,16,0) -- (16,16,500);
      \draw[densely dotted] (112,16,0) -- (112,16,500);
      \draw[densely dotted] (112,112,0) -- (112,112,500);

      \node (P5) at (326,64,500) {};
      \node (P4) at (326,64,400) {};
      \node (P3) at (326,64,300) {};
      \node (P2) at (326,64,200) {};
      \node (P1) at (326,64,100) {};
      \node (P0) at (326,64,0) {};

      \node[right = .7cm of P0] (I) {$\sum$};

      \node[above right = 0.5cm of I] (D0) {\tiny $\mathcal F\n_0 \mathcal F[\m_i(\nb_0\ast\g_{\twi}\ast\V)]$};
      \node[      right = 0.5cm of I] (D1) {\tiny $\mathcal F\n_1 \mathcal F[\m_i(\nb_1\ast\g_{\twi}\ast\V)]$};
      \node[below right = 0.5cm of I] (D2) {\tiny $\mathcal F\n_2 \mathcal F[\m_i(\nb_2\ast\g_{\twi}\ast\V)]$};
      \node[below = 0.2cm of D2] {\tiny $\cdots$};

      \draw[<-] (148,64,500) -- node[fill=white] {\tiny $\mathcal F^{-1}$} (P5);
      \draw[<-] (148,64,400) -- node[fill=white] {\tiny $\mathcal F^{-1}$} (P4);
      \draw[<-] (148,64,300) -- node[fill=white] {\tiny $\mathcal F^{-1}$} (P3);
      \draw[<-] (148,64,200) -- node[fill=white] {\tiny $\mathcal F^{-1}$} (P2);
      \draw[<-] (148,64,100) -- node[fill=white] {\tiny $\mathcal F^{-1}$} (P1);
      \draw[<-] (148,64,0) -- node[fill=white] {\tiny $\mathcal F^{-1}$} (P0);

      \draw[<-] (P5) -- node[right,fill=white] {\tiny $\cdot \mathcal F\g_{\Delta w}$} (P4);
      \draw[<-] (P4) -- node[right,fill=white] {\tiny $\cdot \mathcal F\g_{\Delta w}$} (P3);
      \draw[<-] (P3) -- node[right,fill=white] {\tiny $\cdot \mathcal F\g_{\Delta w}$} (P2);
      \draw[<-] (P2) -- node[right,fill=white] {\tiny $\cdot \mathcal F\g_{\Delta w}$} (P1);
      \draw[<-] (P1) -- node[right,fill=white] {\tiny $\cdot \mathcal F\g_{\Delta w}$} (P0);

      \draw[<-] (P0) -- 
      (I);

      \draw[<-] (I) -- (D0);
      \draw[<-] (I) -- (D1);
      \draw[<-] (I) -- (D2);

    \end{axis}
  \end{tikzpicture}
    \caption{Generating a subgrid $w$-stack ($w$-tower)}
    \label{fig:wtowers}
\end{figure}

\begin{figure}
\centering
  \begin{subfigure}[t]{0.49\columnwidth}
    \centering
  \begin{tikzpicture}[xscale=.8,yscale=.8, inner sep=.5pt]
    \begin{axis}[
        xmin=-0.5,xmax=2.5,ymin=-0.5,ymax=2.5,
        zmin=-2.5,zmax=4,
        width=6cm,height=7cm,
        xlabel=u, ylabel=v, zlabel=w,
        xtick=\empty,ytick=\empty,ztick=\empty,
        axis line style={gray, dashed},
        every axis label/.append style={gray},
        yticklabel pos=right,
      ]


      \pgfplotsinvokeforeach{-2,-1}{
        \addplot3[patch,patch type=rectangle,color=black,fill=white,semitransparent]
        coordinates { (-.25,-.25,#1) (2.25,-.25,#1) (2.25,2.25,#1) (-.25,2.25,#1) };
        \addplot3[patch,patch type=rectangle,color=black,fill=gray,opacity=0.25]
          coordinates { (0,0,#1) (2,0,#1) (2,2,#1) (0,2,#1) };
      }
      \pgfplotsinvokeforeach{0,...,2}{
        \addplot3[patch,patch type=rectangle,color=black,fill=white,semitransparent]
        coordinates { (-.25,-.25,#1) (2.25,-.25,#1) (2.25,2.25,#1) (-.25,2.25,#1) };
        \addplot3[patch,patch type=rectangle,color=black,fill=gray,opacity=0.25]
          coordinates { (0,0,#1) (2,0,#1) (2,2,#1) (0,2,#1) };
        \addplot3+[mesh, solid, no marks, scatter,samples=2, draw=black,
          domain=0.375:0.875, y domain=0.1:0.6, semitransparent]
          { #1 };

        \draw[color=gray] (0.375,0.1,#1) -- (0.375,0.1,#1+1);
        \draw[color=gray] (0.875,0.1,#1) -- (0.875,0.1,#1+1);
        \draw[color=gray] (0.375,0.6,#1) -- (0.375,0.6,#1+1);
        \draw[color=gray] (0.875,0.6,#1) -- (0.875,0.6,#1+1);

      }
      \foreach \w in {3} {
        \draw[<->] (-.25,2.5,3) -- node[inner sep=0, fill=white] {$2\xm$} (2.25,2.5,3);
        \addplot3[patch,patch type=rectangle,color=black,fill=white,semitransparent]
        coordinates { (-.25,-.25,\w) (2.25,-.25,\w) (2.25,2.25,\w) (-.25,2.25,\w) };
        \draw[<->] (0,1.75,3) -- node[inner sep=0, fill=white] {$2\xA$} (2,1.75,3);
        \draw[<->] (0.375,0.7,3) -- node[above right,xshift=-.1cm] {$2\xx_\text{r}$} (0.875,0.7,3);
        \addplot3[patch,patch type=rectangle,color=black,fill=gray,opacity=0.25]
          coordinates { (0,0,\w) (2,0,\w) (2,2,\w) (0,2,\w) };
        \addplot3+[mesh, solid, no marks, scatter,samples=2, draw=black,
          domain=0.375:0.875, y domain=0.1:0.6, semitransparent]
                  { \w };
      }

      \draw[<->] (2.25,2.25,2) -- node[inner sep=-1, fill=white] {$\Delta w$} (2.25,2.25,1);
      \draw[<->] (0.375,-0.25,3) -- node[left] {$2w_\text{r}$} (0.375,-0.25,0);

      \addplot3+[scatter, mark=ball, mark size=1pt,
        scatter/use mapped color={
          color=transparent, ball color=black}] coordinates
                { (0.625, 0.3, 1.5) };

    \end{axis}
    \end{tikzpicture}
  \caption{Spatial-frequency space layout}
  \label{fig:spacings_grid}
  \end{subfigure}
  \begin{subfigure}[t]{0.49\columnwidth}
    \centering
  \begin{tikzpicture}[xscale=.8,yscale=.8, inner sep=0]
    \begin{scope}[xshift=5.5cm]
      \draw[<->] (0,4.75) -- node[inner sep=0, fill=white] {$2\yV$} (5,4.75);
      \draw[dotted] (0,0) rectangle (5,5);

      \draw[fill=gray!25!white] (1,1) rectangle (4,4);
      \foreach \i in { 2, 3 } {
        \draw (\i,1) -- (\i,4);
        \draw (1,\i) -- (4,\i);
      }
      \foreach \i in { 1, 2, 3 } {
        \foreach \j in { 1, 2, 3 } {
          \draw[densely dotted] (\i-0.25,\j-0.25) rectangle (\i+1.25,\j+1.25);
        }
      }
      \draw[<->] (1,3.5) -- node[fill=gray!25!white] {$2\yy_\text{FoV}$} (4,3.5);
      \draw[<->] (0.75,2.5) -- node[fill=gray!25!white] {$2\yn$} (2.25,2.5);
      \draw[<->] (1,1.5) -- node[fill=gray!25!white] {$2\yB$} (2,1.5);


      \draw[<->] (0,4.75) -- node[inner sep=0, fill=white] {$2\yV$} (5,4.75);

    \end{scope}
    \end{tikzpicture}

  \caption{Image-space layout}
  \label{fig:spacings_image}
  \end{subfigure}

  \caption{Subgrid and facet spacings for convolutional gridding}
  \label{fig:spacings}
\end{figure}

Using $w$-stacking and $w$-towers we can efficiently generate
visibility values for regularly spaced $uvw$ subgrids, as shown in
\autoref{fig:spacings_grid}.
This is exactly what we need for gridding algorithms like
$w$-projection\,\citep{cornwell2008noncoplanar},
image domain gridding\,\citep{van2018image}, or
$uvw$-\hbox{(de-)gridding} \citep{2022MNRAS.510.4110Y}. These algorithms
allow us to work with irregularly positioned visibilities
as long as we can provide sufficient samples of certain
$2\xx_r \times 2\xx_r \times 2w_r$ environments around
their locations (e.g. dot and box in \autoref{fig:spacings_grid}).

Gridding typically uses spatial-frequency space
convolutions, which means that just as when we split $\B_j = \nb_j
\ast \n_j$ in \autoref{sct:approximation}, it must be cancelled out
by multiplication with a gridding correction function in image space.
Fortunately, our algorithm works well with
image-space factors.  The natural approach is to combine it with
$\nb_j$ to apply it to facets (like $w$-stacking), but theoretically
we could also merge it with $\n_j$ to apply it per subgrid (like
$w$-towers). This means that we can introduce the gridding correction
function into our framework efficiently, possibly even supporting
different gridding algorithms in different grid regions.

In practice, gridding efficiency depends greatly on how finely
the grid is sampled: the more we oversample the spatial-frequency
grid by padding the image or introducing finer $w$-tower storeys,
the smaller the $2\xx_r \times 2\xx_r \times 2w_r$
region we need to sample\,(\citealt{tan1986aperture,ye2020optimal}; also compare \autoref{fig:x0_per_precision}). Our algorithm
allows us to adjust both parameters easily. Due to linearity
we can simply ignore irrelevant facets, and therefore we can freely
increase the image size and therefore spatial-frequency sampling rate $\yV$,
while only covering a given $\yy_\text{FoV}$ area with facets
(see \autoref{fig:spacings_image}).
Extra $w$-tower storeys are also
easily added. On the other hand, either measure will increase computation,
leading to a trade-off against
gridding complexity.

\section{Algorithm}
\label{sct:Implementation}

We assemble the complete algorithms,  both for going from facets
to $w$-towers of subgrids (\autoref{lbl:algorithm_bwd}) and its
dual for going from subgrids  or  $w$-towers back to facets
(\autoref{lbl:algorithm_fwd}). Separable shifts $\delta^{(u)},
\delta^{(v)}$ and $\hat\xx_i, \hat\xxx_i, \hat\yy_j, \hat\yyy_j$ work
analogously to what was shown in  \autoref{sct:shifts} and \autoref{sct:2D}. Apart from
adding $w$-towers, this closely follows \hyperref[fig:algorithm2D]{Fig. \ref*{fig:algorithm2D}}.

Both directions are linear, meaning that we can skip subgrids or $w$-towers
without visibility coverage, just as we can skip facets outside the field of
view, as shown in \autoref{fig:spacings_image}.  Furthermore, all `foreach'
loops can be executed in parallel. This especially applies to $w$-towers, as
we can parallelise per $w$-tower, $w$-tower storey, and/or visibility
chunk. Even the outer $w$-stacking loop can be parallelised to some degree,
memory availability permitting.

\subsection{Dimensions and shifts}
\label{sct:dim_shifts}

Once  the structure of the algorithm has been established, actually making it
work depends entirely on suitable parameter choices.  Fast Fourier
transforms form the backbone of the algorithm, so we start by
looking at their sizes. In
\autoref{lbl:algorithm_bwd} and \ref{lbl:algorithm_fwd} we perform Fourier
transforms at the padded subgrid size $4\xm\yV$
(\autoref{ln:bwd_sg_fft} and \ref{ln:fwd_sg_fft}), padded facet size
$4\xV\yn$ (\autoref{ln:bwd_init_fft1},\ref{ln:bwd_init_fft2} and \ref{ln:fwd_init_fft1},\ref{ln:fwd_init_fft2}),  and finally $4\xm\yn$ (\autoref{ln:bwd_img_fft1},\ref{ln:bwd_img_fft2} and \ref{ln:fwd_img_fft1},\ref{ln:fwd_img_fft2}),
which derives from the previous values. From \autoref{sct:proof} we know
two more things:
Firstly, the total image size $\xV\yV$ must be divisible by both the
  padded subgrid size $\xm\yV$ and padded facet sizes $\xV\yn$;
Secondly, as $\xm\yn = (\xm\yV)(\xV\yn)(\xV\yV)^{-1}$, the product of
  padded subgrid and facet size must be divisible by the image size.
This clearly favours simple factorisations, for example  powers of two, which
are also optimal for fast Fourier transforms.

\LinesNumbered
\SetInd{0.25em}{0.5em}
\SetAlgoHangIndent{1em}
\begin{algorithm}[t]
  \tiny
    \ForEach{$w$-stacking plane $\bar w$}{
      \ForEach{$w$-tower / subgrid mask $\m$}{
        sg\_image := allocate $(4\xm\yV)^2$ zeroes \\
        \ForEach{$w$-tower-storey / subgrid $i$ with $\m_i=\m$ and $\twi = \bar w$}{
          subgrid := allocate $(4\xm\yV)^2$ zeroes \\
          \ForEach{visibility close enough to $\hat\xx_i / \hat\xxx_i / \dwi$}{
            grid visibility to subgrid using conventional method
          }
          sg\_image += $\mathcal F \g_{\bar{w}-\dwi} \cdot \mathcal F(\text{subgrid})$
          \nllabel{ln:bwd_sg_fft}
        }
        \ForEach{facet $j$ of interest}{
          data[$i,j$] := $\mathcal F \n \cdot \text{sg\_image}[$
            $4\xm(\dyj-\yn)\!:\!4\xm(\dyj+\yn), 4\xm(\dyyj-\yn)\!:\!4\xm(\dyyj+\yn)]$
        }
      }
      -- redistribute data between nodes -- \\
      \ForEach{facet $j$ of interest}{
        fct\_buf := allocate $4\xV\yn\times 4\xV\yB$ zeroes
        \nllabel{ln:bwd_var1}        \\
        \ForEach{subgrid column offset $\hat\xx$ at plane $\bar w$}{
          sg\_buf := allocate $4\xm\yn \times 4\xV\yn$ zeroes  \\
          \ForEach{subgrid $i$ with $\dxi = \hat\xx$ and $\bar w_i = \bar w$}{
            sg\_buf[$:,4(\hat\xxx_i\!-\!\xm)\yn\!:\!4(\hat\xxx_i\!+\!\xm)\yn$]
              += $\mathcal F^{(v)^{-1}}(\text{data}[i,j])$
            \nllabel{ln:bwd_img_fft1} \nllabel{ln:bwd_wrap}
          }
          fct\_buf[$4(\hat\xx\!-\!\xm)\yn\!:\!4(\hat\xx\!+\!\xm)\yn,:$]
          += $\mathcal F^{(u)^{-1}}\left(\mathcal F^{(v)}(\text{sg\_buf})[:,-2\xV\yB\!:\!2\xV\yB]\right)$
          \nllabel{ln:bwd_init_fft1} \nllabel{ln:bwd_img_fft2}
        }
        facet[$j$] +=
          $(\mathcal F\nb_j\mathcal \mathcal F\g_{-\bar{w}}\ast\delta_{\dyj}) \cdot
          \mathcal F^{(u)}(\text{fct\_buf})[-2\xV\yB\!:\!2\xV\yB,:]$        \nllabel{ln:bwd_init_fft2}
      }
    }
  \caption{\label{lbl:algorithm_bwd}$w$-towers $\rightarrow$ facets method}
\end{algorithm}

\begin{algorithm}[t]
\tiny
    \ForEach{$w$-stacking plane $\bar w$}{
      \ForEach{non-zero facet $j$}{
        fct\_buf := $\mathcal F^{(u)^{-1}}\text{pad}\big(4\xV\yn\times 4\xV\yB,$ $(\mathcal F\nb_j\mathcal \mathcal F\g_{\bar{w}}\ast\delta_{\dyj}) \cdot \text{facet}[j]\big)$
        \nllabel{ln:fwd_init_fft1} \\
        \ForEach{subgrid column offset $\hat\xx$ at plane $\bar w$}{
          sg\_buf := $\mathcal F^{(v)^{-1}}\text{pad}\big(4\xm\yn\times4\xV\yn,\mathcal F^{(u)}($
            $\text{fct\_buf}[4(\hat\xx\!-\!\xm)\yn\!:\!4(\hat\xx\!+\!\xm)\yn,:])\big)$
          \nllabel{ln:fwd_init_fft2} \nllabel{ln:fwd_img_fft1} \\
          \ForEach{subgrid $i$ with $\hat\xx_i = \hat\xx$ and $\bar w_i = \bar w$}{
            data[$i,j$] :=
            $\mathcal F^{(v)}\big(\text{sg\_buf}[:,4(\hat\xxx_i\!-\!\xm)\yn\!:\!4(\hat\xxx_i\!+\!\xm)\yn]\big)$
            \nllabel{ln:fwd_img_fft2}
          }
        }
      }
      -- redistribute data between nodes -- \\
      \ForEach{$w$-tower / subgrid mask $\m$}{
        sg\_image := allocate $(4\xm\yV)^2$ zeroes \\
        \ForEach{non-zero facet $j$}{
          sg\_image$[4\xm(\dyj-\yn)\!:\!4\xm(\dyj+\yn), 4\xm(\dyyj-\yn)\!:\!4\xm(\dyyj+\yn)$] +=
          $\mathcal F \n \cdot \text{data}[i,j]$ \nllabel{ln:fwd_cut} \\
        }
        \ForEach{$w$-tower-storey / subgrid $i$ with $\m_i=\m$ and $\twi = \bar w$}{
          subgrid := $\mathcal F^{-1}(\mathcal F \g_{\dwi-\bar{w}} \cdot \text{sg\_image})$
            \nllabel{ln:fwd_sg_fft} \\
          \ForEach{visibility close enough to $\hat\xx_i / \hat\xxx_i / \dwi$}{
            degrid visibility from subgrid using conventional method
          }
        }
      }
    }
  \caption{\label{lbl:algorithm_fwd}Facets $\rightarrow$ $w$-towers method (dual)\vspace{-1cm}}
\end{algorithm}

As established in \autoref{sct:shifts}, these factorisations also
restrict permissible subgrid and facet positions
$\dxi$/$\dxxi$/$\dyj$/$\dyyj$. We assume certain base shift
steps $\Delta\xx$ and $\Delta\yy$ such that all $\dxi$/$\dxxi$ and
$\dyj$/$\dyyj$, respectively, are multiples. As we require all
$\dxi\dyj, \dxxi\dyyj\in\mathbb Z$ we can assume $\Delta\xx\Delta\yy =
1$ without loss of generality, and because $4\xV\yV =
(2\Delta\xx\yV)(2\xV\Delta\yy)$ this means that the product of base
subgrid and facet shifts in pixels must equal the image size.

To ensure that $2\xm\Delta\yy, 2\Delta\xx\yn \in
\mathbb Z$ we can derive the following, 
\begin{align}
2\xm\Delta\yy = \frac{2\xm}{\Delta\xx}
= \frac{4\xm\yV}{2\Delta\xx\yV} \in \mathbb Z
\;\text{,}\;
2\Delta\xx\yn = \frac{2\yP}{\Delta\yy}
= \frac{4\xV\yn}{2\xV\Delta\yy} \in \mathbb Z,
\end{align}
so additionally base subgrid and facet shifts must evenly
divide the padded subgrid and padded facet sizes, respectively.

\subsection{Errors and margins}
\label{sct:Parameters}

\begin{figure}[t]
    \centering
    \input{tikz/x0_per_precision.tikz}%
    \caption{Amount of facet space useable per desired base error
       for prolate spheroidal wave function (PSWF)}
    \label{fig:x0_per_precision}
\end{figure}

In addition to  data sizes, the other fundamental parameter is what level of approximation
we are willing to accept. As established in \autoref{sct:errors}, this
is mostly determined by two factors.  
The first is the base error level, which effectively derives from the PSWF parameter
$W=4\xn\yn$. From this we can especially derive $4\xn\yV$, the
minimum subgrid margin size.
The second factor is the base error magnification due to $\nb$, which depends
directly on $\yB\yn^{-1}$ (i.e. how much facet space we are willing to
sacrifice).

We know that the worst-case error magnification happens at the facet
borders, and therefore we just need to determine how large we can make
$\yB$ until the base error multiplied by $|\mathcal F \n(\yB)|^{-1}$
becomes larger than the acceptable error level (compare
\autoref{fig:overview}). This means that for any target error, we  
have a range of PSWF parameters $W$ to choose from, as illustrated in
\autoref{fig:x0_per_precision}.

Facet--subgrid contributions have size $4\xm\yn$, yet effectively
sample only $4\xA\yB$ points of image--frequency space. Therefore, we
can calculate an upper bound on communication efficiency:
\begin{align}
\text{efficiency} =
\frac{\xA\yB}{\xm\yn}
\quad\le \frac{(\xm-\xn)\yB}{\xm\yn}
\end{align}

\noindent
This already shows a fundamental trade-off between facet and
subgrid margins, as shown in \autoref{fig:parameter_search} (top,
dotted). For larger $W$ we have to give up subgrid area, whereas for
small $W$ we lose facet area to compensate for the worse base error.

For a measure of $w$-tower efficiency, we observe that in
\autoref{sct:wt-error} there is to first order a linear dependency
between maximum tower size $|w|$ and the effective margin reserved for
the $w$-term $\xg' = \xgn-\xn$. Therefore, we can use a
$\yy'_\text{max}$-normalised $uvw$ volume as an efficiency measure
subject to $\xA \le \xm-\xn-\xg'$:
\begin{align}
\text{$w$-tower efficiency} =
\frac{\xg'}{\xm}\left(\frac{\xA\yB}{\xm\yn}\right)^2
\quad\le \frac{ 4(\xm-\xn)^3\yB^2}{27\xm^3\yn^2}.
\end{align}

\noindent Optimally $\xg' \approx \frac13
(\xm-\xn)$, which again yields an upper bound, as shown in
\autoref{fig:parameter_search} (bottom, dotted). As should be
expected, larger $W$ are somewhat less efficient here as   the
PSWF and $\g_w$  compete for limited subgrid space.

\begin{figure}
    \centering
\begin{tikzpicture}

\definecolor{white}{rgb}{1.0,1.0,1.0};
\definecolor{crimson2143940}{RGB}{214,39,40}
\definecolor{darkgray176}{RGB}{176,176,176}
\definecolor{darkorange25512714}{RGB}{255,127,14}
\definecolor{forestgreen4416044}{RGB}{44,160,44}
\definecolor{lightgray204}{RGB}{204,204,204}
\definecolor{mediumpurple148103189}{RGB}{148,103,189}
\definecolor{steelblue31119180}{RGB}{31,119,180}

\begin{groupplot}[
    group style={group size=1 by 2,
      vertical sep=0.15cm,
      xticklabels at=edge bottom
  }]
\nextgroupplot[
height=0.389\columnwidth,
legend cell align={left},
    legend style={ font=\tiny, draw=none, fill=none, at={(0.5,1.03)},
      /tikz/every even column/.append style={column sep=0.0cm},
      anchor=south, legend columns=-1},
tick align=outside,
tick pos=left,
width=.9\columnwidth,
x grid style={darkgray176},
xmin=4, xmax=22,
xtick style={color=black},
y grid style={darkgray176},
ylabel={\tiny$\frac{\xA\yB}{\xm\yn}$},
ymin=1e-06, ymax=0.941439912178207,
ytick style={color=black},
    unbounded coords=discard
]

\addplot [white] coordinates { (0, 0) };
\addlegendentry{\phantom{$^0$}error:}

\addplot [semithick, steelblue31119180, densely dotted, forget plot] table {tikz/effective-8k-2k-1k-000.dat};
\addplot [semithick, steelblue31119180] table {tikz/effective-8k-2k-1k-001.dat};
\addlegendentry{$10^{-3}$}
\addplot [semithick, steelblue31119180, mark=*, mark size=1, mark options={solid}, forget plot]
table {%
9.8125 0.76171875
};
\addplot [semithick, darkorange25512714, densely dotted, forget plot]
table {tikz/effective-8k-2k-1k-002.dat};
\addplot [semithick, darkorange25512714]
table {tikz/effective-8k-2k-1k-003.dat};
\addlegendentry{$10^{-4}$}
\addplot [semithick, darkorange25512714, mark=*, mark size=1, mark options={solid}, forget plot]
table {%
10.875 0.7109375
};
\addplot [semithick, forestgreen4416044, densely dotted, forget plot]
table {tikz/effective-8k-2k-1k-004.dat};
\addplot [semithick, forestgreen4416044]
table {tikz/effective-8k-2k-1k-005.dat};
\addlegendentry{$10^{-5}$}
\addplot [semithick, forestgreen4416044, mark=*, mark size=1, mark options={solid}, forget plot]
table {%
13.5625 0.7109375
};
\addplot [semithick, crimson2143940, densely dotted, forget plot]
table {tikz/effective-8k-2k-1k-006.dat};
\addplot [semithick, crimson2143940]
table {tikz/effective-8k-2k-1k-007.dat};
\addlegendentry{$10^{-6}$}
\addplot [semithick, crimson2143940, mark=*, mark size=1, mark options={solid}, forget plot]
table {%
16.1875 0.7109375
};
\addplot [semithick, mediumpurple148103189, densely dotted, forget plot]
table {tikz/effective-8k-2k-1k-008.dat};
\addplot [semithick, mediumpurple148103189]
table {tikz/effective-8k-2k-1k-009.dat};
\addlegendentry{$10^{-7}$\phantom{error:}}
\addplot [semithick, mediumpurple148103189, mark=*, mark size=1, mark options={solid}, forget plot]
table {%
18.875 0.7109375
};

\nextgroupplot[
height=0.389\columnwidth,
tick align=outside,
tick pos=left,
width=.9\columnwidth,
x grid style={darkgray176},
xlabel={\tiny $W = 4\xn\yn$},
xmin=4, xmax=22,
xtick style={color=black},
y grid style={darkgray176},
ylabel={\tiny
  $\frac{\xg}{\xm}\big(\frac{\xA\yB}{\xm\yn}\big)^2$},
ymin=1e-06, ymax=0.116902420725556,
ytick style={color=black},
yticklabel style={/pgf/number format/fixed, /pgf/number format/precision=2},
scaled y ticks=false
]
\addplot [semithick, steelblue31119180, densely dotted] table {tikz/effective-8k-2k-1k-010.dat};
\addplot [semithick, steelblue31119180] table {tikz/effective-8k-2k-1k-011.dat};
\addplot [semithick, steelblue31119180, mark=*, mark size=1, mark options={solid}]
table {%
8.1875 0.0884551554918289
};
\addplot [semithick, darkorange25512714, densely dotted] table {tikz/effective-8k-2k-1k-012.dat};
\addplot [semithick, darkorange25512714] table {tikz/effective-8k-2k-1k-013.dat};
\addplot [semithick, darkorange25512714, mark=*, mark size=1, mark options={solid}]
table {%
10.875 0.0857479870319366
};
\addplot [semithick, forestgreen4416044, densely dotted] table {tikz/effective-8k-2k-1k-014.dat};
\addplot [semithick, forestgreen4416044] table {tikz/effective-8k-2k-1k-015.dat};
\addplot [semithick, forestgreen4416044, mark=*, mark size=1, mark options={solid}]
table {%
13.5625 0.0830408185720444
};
\addplot [semithick, crimson2143940, densely dotted] table {tikz/effective-8k-2k-1k-016.dat};
\addplot [semithick, crimson2143940] table {tikz/effective-8k-2k-1k-017.dat};
\addplot [semithick, crimson2143940, mark=*, mark size=1, mark options={solid}]
table {%
16.1875 0.0803966075181961
};
\addplot [semithick, mediumpurple148103189, densely dotted] table {tikz/effective-8k-2k-1k-018.dat};
\addplot [semithick, mediumpurple148103189] table {tikz/effective-8k-2k-1k-019.dat};
\addplot [semithick, mediumpurple148103189, mark=*, mark size=1, mark options={solid}]
table {%
18.875 0.0776894390583038
};

\end{groupplot}

\end{tikzpicture}%
    \caption{Parameter efficiencies depending on PSWF and precision for
      \(4\xV\yV=8192, 4\xV\yn=2048, \xm\yV=1024, 4\xV\yy_\text{fov} = 6656\)
      (top: efficiency; bottom: $w$-tower efficiency;
      dotted: upper efficiency bound; marked: best found parameter combination)}
    \label{fig:parameter_search}
\end{figure}

\subsection{Parameter search}

Unfortunately, the side conditions from \autoref{sct:dim_shifts} mean
that at best we can get close to those limits, especially when we try
to realise facet and subgrid partitioning as shown in
\autoref{fig:spacings}:
To account for subgrid or facet overlaps due to offset
  restrictions (see \autoref{sct:shifts}) for the purpose
  of efficiency calculation, $2\xA$ and $2\yB$ should be
  multiples of $\Delta\xx$ and $\Delta\yy$, respectively. As we require the same of $2\xm$ and
  $2\yn$, respectively, this extends to the subgrid and facet margins
  $\xm-\xA$ and $\yn-\yB$.
  
We would also like the facets to cover the field of view
  efficiently. If we need $k\times k$ facets to cover the field of
  view, this means that on average every facet only contains $2\yB' =
  2\yy_\text{fov}k^{-1}$ useable data. Even numbers of facets are a
  special case, as with a centred field of view this requires placing
  facets at an offset of $\yB$. Therefore in this case $\yB$ must be a multiple of $\Delta\yy$ as well.
  
In practice, finding good parameters boils down to trying all combinations
of $W=4\xn\yn$, $\Delta\xx$ and  $\Delta\yy$, $\xA$, and $\yB$
exhaustively. After that, the smallest possible $\yn$ and subsequently
all remaining parameters and efficiency
measures can be determined. If we plot the best solution of an example
parameter search for every value of $W$
(\autoref{fig:parameter_search}, top solid), we see that the
achievable efficiency can be quite unpredictable.  For $w$-towers we
gain another degree of freedom in the $w$-term margin $\xg'$, which
makes efficiency better behaved (\autoref{fig:parameter_search},
bottom solid). This is mainly because the requirement that $(\xA-\xm)$
must be a multiple of $\Delta\xx$ is much less restrictive when we can
still derive efficiency from extra subgrid space by making $w$-towers
taller.

\begin{table}[t]
  \caption{\label{tbl:configs} Sample of possible algorithm parameters
    for a fixed image size (columns $W=4\xn\yn$, $\Delta\xx$ \&
    $\Delta\yy$, $\xA$, and $\yB$) and the corresponding efficiencies
    (last two columns).}
    \centering \tiny \setlength{\tabcolsep}{5pt}
    \begin{tabular}{r|rrr|rr|rr} \hline\hline
      PSWF & \multicolumn{3}{c|}{Facet} & \multicolumn{2}{c|}{Subgrid} & efficiency & $w$-towers\,\\
            $4\xn\yn$ & $4\xV\yn$ & $4\xV\yB$ & $k$ & $4\xm\yV$ & $4\xA\yV$ & & efficiency\\\hline
      13.56 &  2\,048&   1\,664 &  4  & 1\,024 &   896 & 71.1\% & 8.3\% \\
      13.56 &  4\,096&   3\,328 &  2  &    512 &   448 & 71.1\% & 8.3\% \\
      13.56 &  8\,192&   6\,656 &  1  &    256 &   224 & 71.1\% & 8.3\% \\
      21.38 &  1\,792&   1\,664 &  4  & 1\,024 &   640 & 58.0\% & 9.4\% \\
      21.38 &  3\,584&   3\,328 &  2  &    512 &   320 & 58.0\% & 9.4\% \\
      21.38 &  7\,168&   6\,656 &  1  &    256 &   160 & 58.0\% & 9.4\% \\
      16.31 &  1\,536&   1\,344 &  5  & 1\,024 &   640 & 54.2\% & 8.5\% \\
      11.94 &  3\,072&   2\,304 &  3  &    512 &   448 & 63.2\% & 6.4\% \\
       9.75 &  6\,144&   3\,584 &  2  &    256 &   224 & 47.4\% & 3.7\% \\
      18.12 &  1\,280&   1\,152 &  6  & 1\,024 &   640 & 54.2\% & 7.7\% \\
      18.12 &  2\,560&   2\,304 &  3  &    512 &   416 & 70.4\% & 7.7\% \\
      11.06 &  5\,120&   3\,584 &  2  &    256 &   224 & 56.9\% & 5.0\% \\
      13.56 &  1\,024&      832 &  8  & 1\,024 &   512 & 40.6\% & 6.5\% \\
      13.56 &  2\,048&   1\,664 &  4  &    512 &   256 & 40.6\% & 6.5\% \\
      13.56 &  4\,096&   3\,328 &  2  &    256 &   128 & 40.6\% & 6.5\% \\
      \hline
    \end{tabular}
    \vspace{.1cm}

    (image size $4\xV\yV = 8\,192$, field of view $4\xV\yy_\text{fov} = 6\,656$, target error $10^{-5}$)
\end{table}

\autoref{tbl:configs} shows a number of algorithm parameter
  sets for a fixed image size
$4\xV\yV = 8\,192$, optimised for $w$-towers efficiency at $10^{-5}$
target error.  Configurations
with matching contribution sizes $4\xm\yn$ generally have the same
properties as they arrive at the same balance between base
  error (i.e. PSWF parameter $W=\xn\yn$) and error magnification
  (i.e. facet margin $\yB\yn^{-1}$) independently of the concrete values
  of $\yn$ or $\xm$ (compare \autoref{sct:Parameters}). This works as
  long as we can find equivalent facet splits
  (and base offsets) for them.
For instance, in \autoref{tbl:configs} $4\xV\yn = 1280, 4\xm\yV=1024$ is in
  the same family as $4\xV\yn = 5120, 4\xm\yV=256$, but as a facet split of
  $k=1.5$ would be meaningless a different PSWF parameter $W$ must be used.
  Following the same logic we can also scale the image size $\yV$ or resolution
$\xV$, so the configuration from
\autoref{tbl:configs} with facet size
  $4\xV\yn=1024$ could be scaled into a configuration
with $4\xV\yV=524\,288$ (4.4 TB of data) and $4\xV\yn=65\,536$ (69 GB
of data), which is where a  $k\times k=64$-way facet split would
start to seem more appropriate.

\section{Results}
\label{sct:Results}

\begin{table*}[t]
  \caption{\label{tbl:test_configs} Test configuration parameters and results
    (left: scaling baseline length $\xV$; right: scaling field of view (FoV) size  $\yV$)}
  \centering \tiny
  \setlength{\tabcolsep}{5pt}
    \begin{tabular}{rrrr||rrrrr||rrrrr} \hline\hline
      \multicolumn{2}{c}{Image size} & \multicolumn{2}{c||}{Facet size}
        & FoV & Data & \multicolumn{2}{c}{FFT work} & Accuracy
        & FoV & Data & \multicolumn{2}{c}{FFT work} & Accuracy\phantom{\^I}\\
      padded&effective&padded&effective&&& facets & $w$-towers &&&& facets & $w$-towers \\
      $4\xV\yV$ & $4k\xV\yB$          & $4\xV\yn$ & $4\xV\yB$
        & $2\yy_\text{fov}$ & [GB] & [Gflop] & [Gflop] & RMSE
      & $2\yy_\text{fov}$ & [GB] & [Gflop] & [Gflop] & RMSE\\\hline
196\,608 & 159\,744 & 49\,152 & 39\,936 & 0.322 & 997.0 & 131273.7 & 1137193.0 &  9.80$\cdot10^{-6}$ & 0.322  & 997.0  & 131273.8& 1137196.6& 9.80$\cdot10^{-6}$  \\
163\,840 & 133\,120 & 40\,960 & 33\,280 & 0.322 & 982.6 & 81939.9  & 1123209.5 &  9.90$\cdot10^{-6}$ & 0.268  & 561.0  & 63325.4 & 616494.7& 1.15$\cdot10^{-5}$  \\
131\,072 & 106\,496 & 32\,768 & 26\,624 & 0.322 & 918.9 & 45456.9  & 1066548.9 &  9.64$\cdot10^{-6}$ & 0.215  & 280.7  & 26406.9 & 293813.3& 1.29$\cdot10^{-5}$  \\
114\,688 & 93\,176  & 28\,672 & 23\,296 & 0.322 & 826.1 & 31014.4  & 988129.3 &  9.33$\cdot10^{-6}$ & 0.188  & 187.7  & 15267.7 & 190949.7& 1.35$\cdot10^{-5}$  \\
 98\,304 & 79\,872  & 24\,576 & 19\,968 & 0.322 & 678.9 & 20192.4  & 860429.7 &  9.93$\cdot10^{-6}$ & 0.161  & 119.9  & 8900.1  & 116643.0& 1.54$\cdot10^{-5}$  \\
 65\,536 & 53\,248  & 16\,384 & 13\,312 & 0.322 & 318.9 & 6952.6   & 428050.5 &  1.08$\cdot10^{-5}$ & 0.107  & 39.8   & 2129.9  & 33158.6 & 1.73$\cdot10^{-5}$  \\
 49\,152 & 39\,936  & 12\,288 & 9\,984  & 0.322 & 163.8 & 2985.8   & 239863.2 &  1.24$\cdot10^{-5}$ & 0.080  & 19.2   & 773.2   & 13677.3 & 1.97$\cdot10^{-5}$  \\
 32\,768 & 26\,624  & 8\,192  & 6\,656  & 0.322 & 63.9  & 904.0    & 102354.0 &  1.29$\cdot10^{-5}$ & 0.054  & 7.7    & 138.6   & 4795.3  & 2.00$\cdot10^{-5}$  \\
 24\,576 & 19\,968  & 6\,144  & 4\,992  & 0.322 & 32.5  & 413.8    & 54793.0  &  1.45$\cdot10^{-5}$ & 0.040  & 5.5    & 76.2    & 2693.6  & 2.17$\cdot10^{-5}$  \\
 16\,384 & 13\,312  & 4\,096  & 3\,328  & 0.322 & 17.5  & 161.9    & 33241.0  &  1.54$\cdot10^{-5}$ & 0.027  & 3.7    & 34.4    & 1436.9  & 2.34$\cdot10^{-5}$  \\
 12\,288 & 9\,984   & 3\,072  & 2\,496  & 0.322 & 10.4  & 63.4     & 20334.2  &  1.57$\cdot10^{-5}$ & 0.020  & 2.7    & 19.6    & 1129.9  & 2.46$\cdot10^{-5}$ \\
  8\,192 & 6\,656   & 2\,048  & 1\,664  & 0.322 & 7.5   & 29.9     & 13379.6  &  1.64$\cdot10^{-5}$ & 0.013  & 2.0    & 8.3     & 855.2    & 2.58$\cdot10^{-5}$ \\
\hline
    \end{tabular}

    \vspace{.1cm} All using $W = 4\xn\yn = 13.5625$, facets $k=4$, subgrid sizes $4\xm\yV=1024$ padded, $4\xA\yV = 896$ effective.
    `Data': all contribution terms exchanged;
    `FFT work': operations spent on facet--$w$-tower FFTs;
    `Accuracy': difference between degridded visibilities and DFT;
    `Time': wall-clock from first facet to last visibility.
\end{table*}

We investigate scaling properties by posing an SKA-sized `forward'
imaging problem: predicting visibilities for the SKA1-Mid layout with 197
dishes, assuming a snapshot duration of 29 minutes, a dump time of
0.142\,s (12288 dumps), a frequency window of 350\,MHz-472.5\,MHz split into 11264
channels, resulting in $2.6 \cdot 10^{12}$ visibilities (or 42.3\,TB
data at double precision).  The $uv$-coverage for such an observation
is shown in \autoref{fig:layout}.  The phase centre was chosen at a
declination of 0 with hour angles centred around transit (i.e. 0
hour angle). This means around 60 degrees elevation for SKA Mid, which
we can compensate for with shear factor $h_v = 0.5936$ (see
\autoref{sct:w-snapshots}). For the de-gridding we used a
$8\times8\times4$ deconvolution using Sze Meng Tan's
kernels\,\citep{ye2020optimal}, optimised for $x_0=0.40625$
(i.e. $1\!-\!2\!\times\!0.40625\!=$18.75\% image margin) and $x_0=0.1875$, respectively, resulting
in $\sim\!\!10^{-5}$ expected degridding accuracy. The test image
contained ten on-grid sources with the same intensity, all placed
randomly on facet borders to maximise errors, as explained in
\autoref{sct:errors}. To verify correctness, we randomly sampled about
$6 \cdot 10^{8}$ visibilities per run and compared them against a direct
evaluation of the measurement equation.

To make configurations comparable, all tests were conducted with scaled
variants of the first configuration from \autoref{tbl:configs}. We specifically
held facet count and subgrid size constant, but scaled both image and facet size
as well as subgrid count. We configured our
implementation\,\citep{DistributedPredictIO2019,Wortmann2023} so it distributed
work across 16 nodes and 49 processes (1 scheduler, $4\times4=16$ facet workers
and 32 $w$-towers / subgrid workers). The tests were run on 16 dual
Intel\textregistered{} Xeon\textregistered{} Platinum 8368Q (2 $\times$ 76
cores) with 256\,GiB RAM each.

\subsection{Scaling resolution}

The image size in pixels scales as the product of the field of view (and
therefore $\yV$) and the resolution $\xV$. For this section we consider the
effects of scaling $\xV$, while holding the field of view at $2\yy_\text{fov} =
0.322$ ($\sim$\,18.5 degrees;  see middle column of
\autoref{tbl:test_configs}). This corresponds to roughly the third null of SKA1 Mid's beam at the
finest tested image resolution of $4\xV\yV = 196\,608$.  For smaller grid sizes,
this results in coarser image resolution and truncation of the $uv$-coverage
(\autoref{fig:layout}).  This does not decrease degridding work
substantially due to the high number of short baselines (\autoref{fig:scaling},
solid red line).

\begin{figure}
    \tikzset{external/export=false}
    \begin{tikzpicture}
      \definecolor{color0}{rgb}{0.12156862745098,0.466666666666667,0.705882352941177}
      \definecolor{color1}{rgb}{1,0.498039215686275,0.0549019607843137}
      \definecolor{color2}{rgb}{0.172549019607843,0.627450980392157,0.172549019607843}
      \definecolor{color3}{rgb}{0.83921568627451,0.152941176470588,0.156862745098039}

      \begin{loglogaxis}[xlabel={\tiny Image size ($4\xV\yV$)}, ylabel={\tiny Data [byte]\quad Work [flop]},
          x grid style={white!69.0196078431373!black},
          y grid style={white!69.0196078431373!black},
          height=0.7\columnwidth, width=\columnwidth,
          mark size=2pt,
        ]
        \coordinate (legend) at (axis description cs:0.45,0.97);
        \pgfplotstableread[col sep=&, row sep=crcr]{
          image & yb
          & fovb & transb & facetb & subgridb & timeb
          & fova & transa & faceta & subgrida & timea
          & degrida & degridb \\
  8192 & 1664  & 0.322 & 7.54  & 29.9     & 13379.621  &   403.46  & 0.013  & 2.0    & 8.3     & 855.22    &  524.43  & 2108341.11 & 2928712.2\\
 12288 & 2496  & 0.322 & 10.42 & 63.4     & 20334.198  &   390.09  & 0.020  & 2.7    & 19.6    & 1129.945  &  502.79  & 2218326.81 & 2928712.2\\
 16384 & 3328  & 0.322 & 17.51 & 161.9    & 33241.012  &   397.02  & 0.027  & 3.7    & 34.4    & 1436.864  &  518.43  & 2338058.1  & 2928712.2\\
 24576 & 4992  & 0.322 & 32.53 & 413.8    & 54793.026  &   394.18  & 0.040  & 5.5    & 76.2    & 2693.581  &  489.12  & 2435253.5  & 2928712.2\\
 32768 & 6656  & 0.322 & 63.88 & 904.0    & 102354.019 &   417.94  & 0.054  & 7.7    & 138.6   & 4795.348  &  473.84  & 2555950.41 & 2928712.2\\
 49152 & 9984  & 0.322 & 163.8 & 2985.8   & 239863.228 &   480.04  & 0.080  & 19.2   & 773.2   & 13677.308 &  453.95  & 2702734.63 & 2928711.6\\
 65536 & 13312 & 0.322 & 318.9 & 6952.6   & 428050.532 &   608.38  & 0.107  & 39.8   & 2129.9  & 33158.594 &  466.24  & 2798410.96 & 2928712.2\\
 98304 & 19968 & 0.322 & 678.9 & 20192.4  & 860429.694 &   917.76  & 0.161  & 119.9  & 8900.1  & 116642.965&  445.22  & 2922273.7  & 2928712.2\\
114688 & 23296 & 0.322 & 826.1 & 31014.4  & 988129.263 &   979.20  & 0.188  & 187.7  & 15267.7 & 190949.675&  497.13  & 2925991.13 & 2928711.6\\
131072 & 26624 & 0.322 & 918.9 & 45456.9  & 1066548.91 &   1039.30 & 0.215  & 280.7  & 26406.9 & 293813.302&  548.46  & 2927725.98 & 2928644.8\\
163840 & 33280 & 0.322 & 982.6 & 81939.9  & 1123209.45 &   1093.81 & 0.268  & 561.0  & 63325.4 & 616494.742&  765.14  & 2928558.15 & 2928711.8\\
196608 & 39936 & 0.322 & 997.0 & 131273.7 & 1137193.04 &   1187.24 & 0.322  & 997.0  & 131273.8& 1137196.61&  1156.63 & 2928699.45 & 2928699.5\\
        }\scalingtable;


        \path[draw, densely dotted,color={white!69!black}] (axis cs:1e3,1e2)--(axis cs:1e6,1e11);
        \path[draw, densely dotted,color={white!69!black}] (axis cs:1e3,1e3)--(axis cs:1e6,1e12);
        \path[draw, densely dotted,color={white!69!black}] (axis cs:1e3,1e4)--(axis cs:1e6,1e13);
        \path[draw, densely dotted,color={white!69!black}] (axis cs:1e3,1e5)--(axis cs:1e6,1e14);
        \path[draw, densely dotted,color={white!69!black}] (axis cs:1e3,1e6)--(axis cs:1e6,1e15);
        \path[draw, densely dotted,color={white!69!black}] (axis cs:1e3,1e7)--(axis cs:1e6,1e16);
        \path[draw, densely dotted,color={white!69!black}] (axis cs:1e3,1e8)--(axis cs:1e6,1e17);
        \path[draw, densely dotted,color={white!69!black}] (axis cs:1e3,1e9)--(axis cs:1e6,1e18);
        \path[draw, densely dotted,color={white!69!black}] (axis cs:1e3,1e10)--(axis cs:1e6,1e19);
        \path[draw, densely dotted,color={white!69!black}] (axis cs:1e3,1e11)--(axis cs:1e6,1e20);
        \path[draw, densely dotted,color={white!69!black}] (axis cs:1e3,1e12)--(axis cs:1e6,1e21);
        \node[rotate=23,fill=white,text=gray!50!white, inner sep=1pt,draw=gray!50!white, densely dotted]
        at (axis cs: 7.5e4, 4.2e9) {
          \tiny $\sim(\xV\yV)^3$
        };

        \begin{scope}[thick,mark=o]
        \addplot[color0] table[x=image,y expr=1e9*\thisrow{transb}] \scalingtable;
        \label{plot:transb}
        \addplot[color1] table[x=image,y expr=1e9*\thisrow{facetb}] \scalingtable;
        \label{plot:facetb}
        \addplot[color2] table[x=image,y expr=1e9*\thisrow{subgridb}] \scalingtable;
        \label{plot:subgridb}
        \addplot[color3] table[x=image,y expr=1e9*\thisrow{degridb}] \scalingtable;
        \label{plot:degridb}
        \end{scope}
        \begin{scope}[thick,mark=square,mark options={solid},densely dotted]
        \addplot[color0] table[x=image,y expr=1e9*\thisrow{transa}] \scalingtable;
        \label{plot:transa}
        \addplot[color1] table[x=image,y expr=1e9*\thisrow{faceta}] \scalingtable;
        \label{plot:faceta}
        \addplot[color2] table[x=image,y expr=1e9*\thisrow{subgrida}] \scalingtable;
        \label{plot:subgrida}
        \addplot[color3] table[x=image,y expr=1e9*\thisrow{degrida}] \scalingtable;
        \label{plot:degrida}
        \end{scope}
      \end{loglogaxis}
      \matrix [matrix of nodes, anchor=south, row sep=-5pt] at (legend) {
                           & \tiny Data & \tiny FFT facets & \tiny FFT $w$-towers & \tiny Degrid \\
                           & \tiny [byte] & \tiny [flop] & \tiny [flop] & \tiny [flop] \\
            \tiny Scaling baselines ($\xV$): & \ref*{plot:transb} & \ref*{plot:facetb} & \ref*{plot:subgridb} & \ref*{plot:degridb} \\
            \tiny Scaling FoV ($\yV$): & \ref*{plot:transa} & \ref*{plot:faceta} & \ref*{plot:subgrida} & \ref*{plot:degrida} \\
        };

    \end{tikzpicture}
    
  \caption{Complexity scaling (dotted lines indicate cubic growth)}
  \label{fig:scaling}
\end{figure}

On the other hand, higher image resolution increases the cost for facet Fourier
transform work (i.e. \autoref{ln:fwd_init_fft1}, \ref{ln:fwd_img_fft1},
and \ref{ln:fwd_img_fft2} in
\autoref{lbl:algorithm_fwd}) roughly as $O(\xV^2)$ per $w$-stacking plane. For
a baseline distribution where maximum $|w|$ grows linearly with $uv$ distance
(i.e. as long as the Earth's rotation dominates the Earth's curvature)  the number of
$w$-stacking planes required to cover the truncated $uv$-distribution will also
grow linearly with $\xV$. This means the expected worst-case scaling of facet work
is $O(\xV^3)$, and as \autoref{fig:scaling} shows (solid orange line)  we indeed
find scaling slightly better than $O(\xV^3)$.

For $w$-towers, changing image resolution
increases the number of subgrids required to cover all of the baselines.
Therefore, we would expect the growth of subgrid FFT work
(i.e. \autoref{ln:fwd_sg_fft} in \autoref{lbl:algorithm_fwd}) as well
as the data transfer amount to grow with the covered $uvw$ volume, which
increases as $O(\xV^3)$. Compared against
\autoref{fig:scaling} (solid green and blue lines), the scaling is
significantly more efficient. This is clearly because, as first
discussed in \autoref{sct:2D}, we can skip more   $uvw$ volume
the longer the baselines get,  especially as we approach the rare longest
baselines.  By varying subgrid sizes and margins we can trade
$w$-stacking and $w$-tower work, so there is reason to suspect that we will be
able to adapt to bigger telescopes with significantly better-than-cubic net
growth in complexity.

\subsection{Scaling the field of view}
\label{sct:fov_scaling}

\begin{figure*}
  \definecolor{color0}{rgb}{0.12156862745098,0.466666666666667,0.705882352941177}
  \definecolor{color1}{rgb}{1,0.498039215686275,0.0549019607843137}
  \definecolor{color2}{rgb}{0.172549019607843,0.627450980392157,0.172549019607843}
  \definecolor{color3}{rgb}{0.83921568627451,0.152941176470588,0.156862745098039}
  \definecolor{color4}{rgb}{0.580392156862745,0.403921568627451,0.741176470588235}
  \definecolor{color5}{rgb}{0.549019607843137,0.337254901960784,0.294117647058824}
    \tikzset{external/export=false}
    \begin{tikzpicture}

      \begin{axis}[
          axis y line*=right,
          axis x line=none,
          ylabel={\tiny Wall-clock time [s]},
          xmin=0, xmax=1,
          ymin=0, ymax=1439.1447368421, 
          height=0.43\textwidth, width=0.93\textwidth,
        ]
      \end{axis}
      \begin{axis}[
          axis y line*=left,
          xlabel={\tiny Image size ($4\xV\yV$)}, ylabel={\tiny Thread time [s]},
          x grid style={white!69.0196078431373!black},
          y grid style={white!69.0196078431373!black},
          xmin=0, xmax=204804,
          ymin=0, ymax=3.5e6,
          xtick={16384,32768,49152,65536,98304,114688,131072,163840,196608},
          xticklabels={16\,384,32\,768,49\,152,65\,536,98\,304,114\,688,131\,072,163\,840,196\,608},
          ytick={486400,972800,1459200,1945600,2432000,2918400,3404800},
          height=0.43\textwidth, width=0.93\textwidth,
          mark size=2.5pt,
          scaled x ticks=false,
          every y tick scale label/.style={at={(yticklabel cs:0.95))},anchor=south west,outer sep=10pt},
        ]
        \coordinate (legend) at (axis description cs:0.56,0.98);
        \pgfplotstableread[col sep=&, row sep=crcr]{
          image &
          workerfft & workerdegrid & workeridle & facetwork & facetidle & wallclock &
          workerfftb & workerdegridb & workeridleb & facetworkb & facetidleb & wallclockb &
          wallclockd & degridd & fftd & wscreend & indexd \\
  8192 & 3931.4987 & 186801.84 & 328351.87 & 10.8	 & 37.8          & 541217.3  & 309.84    & 260161.37 & 301137.55 & 3.7     & 8.2     & 587595.5   & 10351.138 & 9238.6208  & 177.6009   & 32.5773   & 177.2387 \\
 12288 & 5956.748  & 201236.86 & 312683.2  & 16.9	 & 31.1          & 545132.8  & 435.91    & 263242.46 & 301099.35 & 5.3     & 5.3     & 594332.2   & 11481.462 & 9930.0776  & 423.4732   & 79.3258   & 227.734  \\
 16384 & 9450.734  & 216811.3  & 299838.59 & 46.0	 & 138.6	 & 544281.6  & 516.39    & 261903.02 & 293412.62 & 7.2     & 7.5     & 573417.0   & 13054.596 & 11020.7482 & 784.9115   & 164.8788  & 271.0826 \\
 24576 & 15368.476 & 232161.23 & 280419.85 & 118.7	 & 268.5	 & 554228.5  & 913.89    & 264299.58 & 297362.69 & 13.2    & 7.2     & 575241.0   & 16924.52  & 12601.053  & 2235.9903  & 471.4954  & 479.634  \\
 32768 & 28181.087 & 249736.73 & 258110.23 & 270.5	 & 744.3	 & 581175.0  & 1496.15   & 264421.40 & 296754.37 & 23.1    & 11.8    & 594915.8   & 22860.24  & 13964.3119 & 5307.0294  & 1100.1348 & 920.9899 \\
 49152 & 69455.16  & 283538    & 203321.2  & 1033.3	 & 2273.9	 & 598661.1  & 3916.88   & 269237.38 & 285573.85 & 237.8   & 255.8   & 587960.3   & 49181.89  & 16266.9256 & 21535.6058 & 4543.7658 & 2980.1256\\
 65536 & 138172.33 & 311780.09 & 138275.82 & 2911.0	 & 9002.0	 & 626653.4  & 8886.97   & 273876.69 & 272936.57 & 687.4   & 643.6   & 573781.8   & 174833.91 & 18678.49   & 110584.59  & 21089.88  & 14071.93 \\
 98304 & 325995.99 & 352728.97 & 176469.28 & 10012.5	 & 20947.5	 & 886585.6  & 33168.85  & 288037.35 & 249650.05 & 2886.8  & 2534.4  & 600849.9   & nan & nan & nan & nan & nan \\
114688 & 371472.9  & 351269.39 & 264613.03 & 15566.0	 & 24079.5	 & 1022485.8 & 57060.24  & 299485.37 & 228451.64 & 5059.3  & 3914.0  & 613690.9   & nan & nan & nan & nan & nan \\
131072 & 399543.6  & 356871.61 & 308028.84 & 21540.9	 & 26301.8	 & 1093816.3 & 93303.54  & 314789.43 & 191644.72 & 9680.9  & 8206.2  & 631979.5   & nan & nan & nan & nan & nan \\
163840 & 409311.6  & 354931.2  & 494686.7  & 33656.4	 & 25520.9	 & 1299320.3 & 210164.09 & 330318.66 & 293731.67 & 24040.1 & 15384.5 & 869634.6   & nan & nan & nan & nan & nan \\
196608 & 399214.1  & 345946.77 & 830516.8  & 50285.8	 & 24438.7	 & 1629513.0 & 399065.50 & 346044.80 & 836839.20 & 50102.8 & 23698.7 & 1626764.8  & nan & nan & nan & nan & nan \\
        }\timingtable;


        \begin{scope}[thick,mark=o,mark options={solid},densely dotted]
        \addplot[color1] table[x=image,y expr=\thisrow{facetworkb}] \timingtable;
        \label{plot:facetworkb}
        \addplot[color2] table[x=image,y expr=\thisrow{workerfftb}] \timingtable;
        \label{plot:workerfftb}
        \addplot[color3] table[x=image,y expr=\thisrow{workerdegridb}] \timingtable;
        \label{plot:workerdegridb}
        \addplot[color4] table[x=image,y expr=\thisrow{workeridleb}] \timingtable;
        \label{plot:workeridleb}
        \addplot[color0] table[x=image,y expr=\thisrow{wallclockb}] \timingtable;
        \label{plot:wallclockb}
        \end{scope}

        \begin{scope}[thick,mark=square]
        \addplot[color1] table[x=image,y expr=\thisrow{facetwork}] \timingtable;
        \label{plot:facetwork}
        \addplot[color2] table[x=image,y expr=\thisrow{workerfft}] \timingtable;
        \label{plot:workerfft}
        \addplot[color3] table[x=image,y expr=\thisrow{workerdegrid}] \timingtable;
        \label{plot:workerdegrid}
        \addplot[color4] table[x=image,y expr=\thisrow{workeridle}] \timingtable;
        \label{plot:workeridle}
        \addplot[color0] table[x=image,y expr=\thisrow{wallclock}] \timingtable;
        \label{plot:wallclock}
        \end{scope}

        \begin{scope}[thick,mark=o,mark options={solid},densely dashed]
        \addplot[color2] table[x=image,y expr=76*2*\thisrow{fftd}] \timingtable;
        \label{plot:workerfftd}
        \addplot[color3] table[x=image,y expr=76*2*\thisrow{degridd}] \timingtable;
        \label{plot:workerdegridd}
        \addplot[color0] table[x=image,y expr=76*2*\thisrow{wallclockd}] \timingtable;
        \label{plot:wallclockd}
        \end{scope}

      \end{axis}

      \matrix [matrix of nodes, anchor=south, row sep=-5pt] (mtx) at (legend) {
                           & \tiny Completion & \tiny FFT facets & \tiny FFT $w$-stacking/towers & \tiny Degrid & \tiny Idle & \quad\quad & \tiny measurement method / conversion\\
        \tiny SwiFTly, scaling baselines ($\xV$):
        & \ref*{plot:wallclock} \tiny[w] & \ref*{plot:facetwork} \tiny[t] & \ref*{plot:workerfft} \tiny[t] & \ref*{plot:workerdegrid} \tiny[t] & \ref*{plot:workeridle} \tiny[t] && \tiny [w] = wall-clock \\
             \tiny SwiFTly, scaling FoV ($\yV$): & \ref*{plot:wallclockb} \tiny[w] & \ref*{plot:facetworkb} \tiny[t] & \ref*{plot:workerfftb} \tiny[t] & \ref*{plot:workerdegridb} \tiny[t] & \ref*{plot:workeridleb} \tiny[t] && \tiny [c] = per $16\times2$ CPUs \\
             \tiny \texttt{ducc}, scaling FoV ($\yV$): & \ref*{plot:wallclockd} \tiny[c] & & \ref*{plot:workerfftd} \tiny[c] & \ref*{plot:workerdegridd} \tiny[c] &&& \tiny [t] = per $16\times2\times2\times76$ threads\\
        };

    \end{tikzpicture}

    \caption{Benchmark results.
      \texttt{ducc} results scaled as if distributed perfectly to 16 nodes}
  \label{fig:timing}
\end{figure*}

To investigate field of view scaling, let us now hold the grid dimension $\xV$
constant such that it fits all baselines from \autoref{fig:layout}
exactly. Changing image size $4\xV\yV$ then will change $\yV$, and therefore the
effective field of view size $2\yy_\text{fov}$. In our test scenarios shown
in \autoref{tbl:test_configs} we again scale   from the maximum case of 16.8
degrees down to 0.7 degrees.

The reasoning about computational complexity is more complex in
this case.  To determine the number of $w$-stacking planes, we first note that
scaling $\yV$ means scaling frequency-space resolution; therefore, a given
subgrid size $4\xm\yV$ and its margins $4(\xm-\xA)\yV$ now cover less
frequency space. This means that for the purpose of the discussion in
\autoref{sct:wt-error}, the permissable $\xgn$ decreases, while the relevant
field of view and therefore $\yy_\text{max}$ increases. It follows that the
number of required $w$-stacking planes increases as $O(\yV^2)$ to first order. As each plane is also $O(\yV^2)$ larger,
the total facet Fourier transform complexity growth is about $O(\yV^4)$.
The cost for subgrid Fourier transforms will simply
increase with the number of subgrids we need to produce. As the
subgrids cover less spatial-frequency space each, we need to increase their number by $O(\yV^2)$
to cover the telescope layout in $uv$. However, to implement (de-)gridding (e.g. \cite{2022MNRAS.510.4110Y}) we also need to decrease the $w$ distance between
$w$-tower planes as the image curvature $n \sim \yy_\text{fov}^2$
increases. The result is an overall scaling of $O(\yV^4)$
again. Interestingly enough, this means that the number
of storeys in a $w$-tower is constant. The tower layout shrinks linearly
along the $u$- and $v$-axes, and about quadratically along the $w$-axis.

As \autoref{fig:scaling} shows, the experiments again demonstrate slightly
better than expected scaling behaviour; facet work   grows super-cubically
once we get to larger image sizes, but the subgrid Fourier transform work and
data transfer volume stay at or slightly below $O(\yV^3)$. This is because a
larger number of fine-grained subgrids and $w$-towers allow us to more closely
match the baseline distribution. In a way we are observing the fractal
dimension of the telescope layout.   We note that existing  widefield imaging
algorithms would show the same $O(\yV^4)$ scaling (e.g. using
constant-size facets would see a $O(\yV^2)$ growth both in the number of
visibilities and facet count for a combined scaling of $O(\yV^4)$ of
phase rotation cost growth).

\subsection{Communication}

In the largest configuration we are dealing with an image of size 196\,608$^2$,
which at double precision translates to about 618.5\,GB of data. To ensure
the accuracy of the gridder we   leave 18.75\,\% margins, which means that the
effective field of view size corresponds to about 408.3\,GB of data. The results in
\autoref{tbl:test_configs} show that 997.0\,GB of communication were exchanged,
so a bit more than twice the size of the information-containing part of the
image.

This is quite efficient, as the configuration has 33 $w$-stacking planes, with
$w$-towers decomposing them into 154 subplanes each, so we are effectively
sampling a $uvw$ grid $33\cdot154\cdot618.5\,\text{GB}=2.6\,\text{EB}$ in
size. We can also contrast against the visibility volume of 42.3\,TB, which
puts the network overhead at about 1.8\,\% relative to the throughput required
to load visibilities from distributed storage.

\subsection{Performance}

\autoref{fig:timing} shows the run-time performance of tests in
\autoref{tbl:test_configs} and compares them to equivalent benchmark
runs of the \texttt{ducc} degridding
kernel\,\citep{arras2021ducc}. The \texttt{ducc} module uses the same
class of degridding algorithm\,\citep{2022MNRAS.510.4110Y} and is in
regular scientific use as a (de)gridding kernel for the
\hbox{WSClean} widefield imager\,\citep{offringa2014wsclean}. For this
test, image data was passed to \texttt{ducc} as one merged
effective image, as in \autoref{tbl:test_configs}, which was
then internally padded according to \texttt{ducc}'s heuristic (target
precision $\varepsilon=10^{-4}$). We predicted visibilities as 598
chunks of 32 baselines each ($\sim$70\,GB). To minimise
redundant FFT work due to repeating $w$-stacking planes, baselines were ordered by
median $w$ before chunking. All chunks were predicted sequentially on
a single CPU (76 cores), with performance linearly extrapolated to 16
dual-CPU nodes for \autoref{fig:timing}.

Our implementation significantly outperforms \texttt{ducc} in raw
degridding speed (red lines) despite the similar degridding
approach. There are a number of possible explanations. Our kernel was
hand-optimised for AVX2, uses a pre-tabulated kernel, and also
convolves across all $w$-tower storeys in one go. Meanwhile,
\texttt{ducc} has a more generic implementation that generates the
convolution kernel on the fly. This calculation even has to be
repeated multiple times per visibility, due to the high memory pressure
of $w$-stacking. Our implementation likely further benefits from
executing degridding and $w$-towers FFTs in parallel as these
workloads have very different operational intensity, and therefore work
well with hyper-threading.

\begin{figure*}[t]
    \centering
    \begin{tikzpicture}

      \begin{axis}[width=1.296\columnwidth,height=0.9\columnwidth,
                   xlabel={\tiny row ($u$)}, ylabel={\tiny column ($v$)},
                   minor ytick={0,...,107}, minor xtick={-107,...,107}, 
                   colorbar, colormap/cool, colorbar style={yticklabel=$10^{\pgfmathprintnumber{\tick}}$},
        ]
        \addplot[only marks, scatter, scatter src=0, point meta=explicit, mark=square*,draw=none, mark size=0.0022\columnwidth]
           table [y expr=\thisrowno{0}-107,
                  x expr=\thisrowno{1}-107,
                  meta expr=log10(\thisrowno{2}*16384), 
                  col sep=comma] {tikz/192-n48k-1k-midr5-chunks.csv};
    \end{axis}
    \end{tikzpicture}
    
    \caption{Visibilities per subgrid position (SKA1-Mid layout, summed over all $w$-stacking planes and $w$-tower storeys).}
    \label{fig:layout}
\end{figure*}

For the traditional $w$-stacking method used by \text{ducc}, FFT work dominates
past images sizes of 32\,768$^2$ due to unmitigated $O(\yV^4)$ complexity
growth (see \autoref{sct:fov_scaling}, green dashed line).  In comparison, the
$w$-towers approach together with the shear transformation from
\autoref{sct:w-snapshots} substantially blunt the impact of larger fields of
view in terms of number and size of required FFTs (green dotted line). Time
spent on FFTs still eventually overtakes degridding despite the lower operation
count (compare \autoref{fig:scaling}) due to lower operational intensity.

Finally, we consider the cost of distribution, represented through the
facet FFT costs (orange lines) and scheduling inefficiencies (violet
lines). The former are basically insignificant, while the latter
eventually start dominating. This is so because while there is an
abundance of parallel tasks, they vary greatly in terms of
computational and communication complexity. This makes
saturating 4864 threads quite challenging in practice.

Our simple scheduling algorithm approaches this by alternating `expensive'
(many visibilities) and `cheap' (few visibilities) $w$-stacking planes to
balance FFT and degridding work as evenly as possible throughout. It then
dynamically schedules subgrid tasks to subgrid workers from a pre-populated work
queue. For the most densely populated subgrids it splits the degridding work
further to allow multiple subgrid workers to work on them in parallel.
Queue sizes were optimised for the large-scale case, and kept constant for all
test runs. This also explains why small configurations sometimes have worse
scheduling efficiency, as due to the small number of subgrids work is scheduled
before imbalances can become apparent. As a result, some workers end up running idle
towards the end of the run. It is clear that effective work scheduling
is one of the main remaining challenges with this algorithm.

\section{Conclusion}
\label{sct:final}

We have presented a scalable way to distribute interferometric
imaging calculations by using spatial distribution and window
functions. This results in a streaming widefield Fourier transform
algorithm that enables interferometry imaging to be parallelised to
thousands of distributed threads, while taking full advantage of both
image and $uvw$ sparseness to reduce transfer and compute costs.

\subsection{Future improvements}

There are still a number of improvements and generalisations that can be made to
the algorithm. Centring contribution terms as in \autoref{sct:shifts} is
not actually necessary, and removing the $\dxi\dyj\in\mathbb Z$ side
condition allows more flexible subgrid and facet placement. Additional
efficiency could be gained by combining different parameters within the same
imaging task (e.g. adjust subgrid size per $w$-level), or even introducing
additional layers between $w$-stacking and $w$-towers on either side of the
distribution pattern (for dealing with very large images). There is good reason
to believe that such techniques could improve performance further.

\subsection{Implementation}

Despite our efforts, the MPI implementation\,\citep{Wortmann2023} is still severely limited by
its primitive work scheduling. A more sophisticated work-stealing mechanism
will likely be required as data flow predictability will only get worse from
here. For instance, \hbox{(de-)gridding} from and to subgrids could be tackled by
accelerators, and storage might struggle to keep pace with visibility data
flows: processing $\sim$40 TB in under 4 min means 10 GB/s/node,
with non-obvious access patterns.

Additionally, to integrate calibration as well as corrections for various
instrumental and environmental effects, our algorithm will have to work in
concert with other state-of-the-art radio astronomy algorithms.  These methods
come with significant complexity of their own, which makes integration even
trickier. We are planning to investigate use of a
specialised execution framework to take over the data transfer and task
scheduling functions of the prototype code.

\subsection{Imaging and calibration}
\label{sct:pipelines_discussion}

Transformations between the sky and aperture planes are common and well-defined
operations in radio-astronomy pipelines, which means that the presented algorithm can help distribute
a variety of functions relevant to radio astronomy. In  \autoref{sct:gridding} we
discuss how it can work with a variety of gridding approaches;   extensions
to cover polarisation, Taylor terms, or point-spread function generation are
straightforward. Furthermore, as every subgrid gets represented in image space along the way, we
can introduce slow-changing image-space multipliers (e.g. corrections for primary beam effects) very cheaply and at good
resolution. This might complement approaches that deal with fast-changing direction-dependent effects such as
IDG\,\citep{van2018image}.

On the other hand, pipelines that require traversing visibilities in a specific ordering
may need substantial rework to make use of the presented algorithm
because much of its efficiency derives from traversing the grid in a
very particular order: iterating over $w$-stacking planes, then over
$u$-columns, and finally visiting individual subgrids. This is not an issue
for a CLEAN major loop iteration that predicts
visibilities to be subtracted for an imaging step; the forward and
backward directions of the algorithm can be combined so that
processing proceeds from a model image to the current residual dirty
image, while touching every observed visibility exactly once.
Combined with a distributed by-facet image-based `minor loop'
deconvolution step, this should yield a scalable CLEAN pipeline.

The presented algorithm is however a worse fit for standard antenna-based
calibration approaches as they require combining data from all baselines to
reliably solve for antenna-based gain terms. The algorithm is designed
such that visibilities from different $w$-stacking planes would basically never
meet in the same memory space. A possible solution would be to reduce the size
of the calibration problem representation enough so that we could make progress even
with an incomplete
view\,\citep[e.g.][]{yatawatta2015distributed}. Alternatively, one might
attempt to devise a distributed algorithm that re-aggregates optimisation
problems, similarly to how this distributed Fourier transform algorithm
re-assembles a complete image.

\subsection{Outlook}

The presented algorithm significantly improves on existing
radio-astronomy imaging algorithms by enabling the  distribution of both
the computation and the working memory load. This enables imaging in
less wall-clock time by distribution over relatively small
general-purpose (and therefore cost effective) nodes. Reducing
wall-clock time to process  data sets is a key requirement for
telescopes like SKA where storage of visibilities during processing is
a major cost driver.

\vfill\break

\begin{acknowledgements}
  Tim Cornwell originally suggested the possibility for such an
  approach at one of the SKA SDP face-to-face meetings, and provided a
  lot of helpful discussion along the way.  The work by Bas van der
  Tol and Bram Veenboer on IDG was a major inspiration, with Bram
  originally suggesting subgrids as a unit of distribution. Steve Gull
  and Sze Meng Tan helped me understand how to perform accurate
  wide-field degridding.  Thanks to Martin Reinecke for helping us
  understand \texttt{ducc}'s performance. We also thank the
    anonymous referee who's comments helped us improve the
    presentation substantially.  Finally we would like to thank the
  SKA SDP Consortium, the SKA Organisation and SKA Observatory for
  providing us the means to do this work.
\end{acknowledgements}

\bibliographystyle{aa}
\bibliography{ska}

\appendix\onecolumn

\section{Reference implementation}

The one-dimensional Fourier transform algorithm can be
demonstrated quite succinctly, as the following snippet shows.
 We note that this is just the base algorithm, without
non-coplanarity corrections.
\vspace{.5cm}

\begin{minipage}{17cm}
\colorlet{codegreen}{green!50!black}
\colorlet{codegray}{blue!50!black}
\colorlet{codepurple}{red!50!black}
\lstset{
  language=Python,
  basicstyle=\fontfamily{pcr}\tiny,
  frame=tb,
  showstringspaces=false,
  commentstyle=\color{codegreen},
  keywordstyle=\color{codepurple},
  numberstyle=\tiny\color{codegray},
  stringstyle=\color{codegray},
  escapeinside={(*}{*)},
  breaklines=false,
  postbreak=\mbox{\textcolor{red}{$\hookrightarrow$}\space}
}
\linespread{1}
\begin{lstlisting}
import numpy as np
import scipy.special
def fft(a): return np.fft.fftshift(np.fft.fft(np.fft.ifftshift(a)))
def ifft(a): return np.fft.fftshift(np.fft.ifft(np.fft.ifftshift(a)))
def pad(ff, N): return np.pad(ff,N//2-ff.shape[0]//2)
def extract(a, N): return a[a.shape[0]//2 - N//2 : a.shape[0]//2 + N//2]

# Set parameters
W = 13.25      # (*\color{codegreen}$= 4\xn\yn$*) (window function size)
N = 8192       # (*\color{codegreen}$= 4\xV\yV$*) (image size)
lB_size = 2048 # (*\color{codegreen}$= 4\xV\yB$*) (facet size)
lN_size = 2560 # (*\color{codegreen}$= 4\xV\yn$*) (padded facet size)
uA_size = 944  # (*\color{codegreen}$= 4\xA\yV$*) (subgrid size)
uM_size = 1024 # (*\color{codegreen}$= 4\xm\yV$*) (padded subgrid size)
uM_lN_size = uM_size * lN_size // N # (*\color{codegreen}$= 4\xm\yn$*) (contribution size)
# Generate constants (prolate-spheroidal)
pswf = scipy.special.pro_ang1(1, 1, np.pi*W/2, 2*np.mgrid[-lN_size//2:lN_size//2] / lN_size)[0]
pswf[0] = 0 # zap NaN
Fb = 1 / extract(pswf, lB_size)                                                           # (*\color{codegreen}$\mathcal F[\nb]$*)
Fn = extract(fft(extract(ifft(pad(pswf, N)), uM_size)), uM_lN_size).real                  # (*\color{codegreen}$\mathcal F[\n]$*)
# Layout subgrids & facets
nsubgrid = (N + uA_size - 1) // uA_size
nfacet = (N + lB_size - 1) // lB_size
sg_off = uA_size * np.arange(nsubgrid)                                                    # (*\color{codegreen}$2\dxi\yV$*)
fct_off = lB_size * np.arange(nfacet)                                                     # (*\color{codegreen}$2\xV\dxxi$*)
sg_alloc = np.roll(np.repeat(np.arange(nsubgrid), uA_size)[:N], N//2-uA_size//2)
fct_alloc = np.roll(np.repeat(np.arange(nsubgrid), lB_size)[:N], N//2-lB_size//2)
sg_A = [ extract(np.roll(sg_alloc, -sg_off[i]), uA_size) == i for i in range(nsubgrid) ]  # (*\color{codegreen}$\A_i$*)
fct_B = [ extract(np.roll(fct_alloc, -fct_off[i]), lB_size) == i for i in range(nfacet) ] # (*\color{codegreen}$\mathcal F[\B_j$]*)
# Generate random grid and image
I = np.random.rand(N)-0.5                                                                 # (*\color{codegreen}$\V$*)
FI = fft(I)                                                                               # (*\color{codegreen}$\mathcal F[\V]$*)
subgrid = [ sg_A[i] * extract(np.roll(I, -sg_off[i]), uA_size) for i in range(nsubgrid) ] # (*\color{codegreen}$\A_i\V$*)
facet = [ fct_B[j] * extract(np.roll(FI, -fct_off[j]), lB_size) for j in range(nfacet) ]  # (*\color{codegreen}$\mathcal F[\B_j\ast\V]$*)

# SwiFTly forwards - facets to subgrids
FmbI = np.empty((nsubgrid, nfacet, uM_lN_size), dtype=complex)
for j in range(nfacet):
    bI = ifft(pad(facet[j] * Fb, lN_size))                                         # (*\color{codegreen}$\mathcal \nb_j\ast\V$*)
    for i in range(nsubgrid):
        FmbI[i,j] = fft(extract(np.roll(bI, -sg_off[i]*lN_size//N), uM_lN_size))   # (*\color{codegreen}$\mathcal F[\m_i(\nb_j\ast\V)]$*)
# - redistribution of (*\color{codegreen}$\mathcal F[\m_i(\nb_j\ast\V)]$*) -
AnmbI = np.empty((nsubgrid, uA_size), dtype=complex)
for i in range(nsubgrid):
    FnmbI = np.zeros(uM_size, dtype=complex)
    for j in range(nfacet):
        FnmbI += np.roll(pad(Fn * FmbI[i,j], uM_size), fct_off[j]*uM_size//N)      # (*\color{codegreen}$\sum_j\mathcal F[\n_j\ast\m_i(\nb_j\ast\V)]$*)
    AnmbI[i] = sg_A[i] * extract(ifft(FnmbI), uA_size)                             # (*\color{codegreen}$\A_i\sum_j(\n_j\ast\m_i(\nb_j\ast\V))$*)
errors = [ AnmbI[i] - subgrid[i] for i in range(nsubgrid) ]                        # ... (*\color{codegreen}$\approx \mathcal \A_i\V$*)
print("RMSE:", np.sqrt(np.mean(np.abs(errors)**2)))

# SwiFTly backwards - subgrids to facets
FnAI = np.empty((nsubgrid, nfacet, uM_lN_size), dtype=complex)
for i in range(nsubgrid):
    FAI = fft(pad(subgrid[i], uM_size))                                            # (*\color{codegreen}$\mathcal F[\A_i\V]$*)
    for j in range(nfacet):
        FnAI[i,j] = Fn * extract(np.roll(FAI, -fct_off[j]*uM_size//N), uM_lN_size) # (*\color{codegreen}$\mathcal F[\n_j\ast\A_i\V]$*)
# - redistribution of (*\color{codegreen}$\mathcal F[\n_j\ast\A_i\V]$*) -
FbmnAI = np.empty((nfacet, lB_size), dtype=complex)
for j in range(nfacet):
    mnAI = np.zeros(lN_size, dtype=complex)
    for i in range(nsubgrid):
        mnAI += np.roll(pad(ifft(FnAI[i,j]), lN_size), sg_off[i]*lN_size//N)       # (*\color{codegreen}$\sum_i\m_i(\n_j\ast\A_i\V)$*)
    FbmnAI[j] = Fb * fct_B[j] * extract(fft(mnAI), lB_size)                        # (*\color{codegreen}$\mathcal F[\nb_j\ast \sum_i\m_i(\n_j\ast\A_i\V)]$*)
errors = [ FbmnAI[j] - facet[j] for j in range(nfacet) ]                           # ... (*\color{codegreen}$\approx \mathcal F[\B_j\ast\V]$*)
print("RMSE:", np.sqrt(np.mean(np.abs(errors)**2)))
\end{lstlisting}
\end{minipage}

\end{document}